%% file: main.tex
\documentclass[aps,prd,twocolumn,amsmath,amssymb,twoside,superscriptaddress,floatfix,nofootinbib,natbib]{revtex4-1}
\usepackage{graphicx}
\usepackage{subfigure}
\usepackage{color}
\usepackage{bm}% bold math  % MM ADDED THIS
\usepackage{amsmath} % MM ADDED THIS
\usepackage{threeparttable}

\newcommand{\comment}[1]{{}}
\def\beq{\begin{equation}}
\def\eeq{\end{equation}}
\def\beqn{\begin{eqnarray}}
\def\eeqn{\end{eqnarray}}

\def\2gcm{\textrm{g cm$^{-2}$}}

\def\planck{{\it Planck}}

\def\H0{\ensuremath{\mathrm{H}_0}}

\def\bl{\bmm{l}}
\def\bL{\bmm{L}}

\newcommand{\bmm}[1]{{\mathbf{#1}}}

\newcommand{\aap}{{Astron.~Astrophys.}}

\newcommand{\jcap}{{J.~Cosm.~Astrop.~Phys.}}

\newcommand{\procspie}{{Proc.~SPIE}}

\newcommand{\mnras}{{Mon.~Not.~R.~Astron.~Soc.}}

\def\mnras{Mon. Not. R. Astron. Soc}
\def\physrep{Physics Reports}

\usepackage{graphicx}
\usepackage{dcolumn}
\usepackage{bm}

\begin{document}
\title{The Atacama Cosmology Telescope: Two-Season ACTPol Lensing Power Spectrum}

\newcommand{\cita}{Canadian Institute for Theoretical Astrophysics, University of Toronto, Toronto, ON, Canada M5S 3H8}

\newcommand{\princetonastro}{Department of Astrophysical Sciences, Peyton Hall, Princeton University, Princeton, NJ 08544}

\newcommand{\dunlap}{Dunlap Institute, University of Toronto, 50 St. George St., Toronto, ON, Canada M5S 3H4}

\newcommand{\ukznastro}{Astrophysics and Cosmology Research Unit, School of Mathematics, Statistics and Computer Science, University of KwaZulu-Natal, Durban 4041, South Africa}

\newcommand{\princetonphysics}{Joseph Henry Laboratories of Physics, Jadwin Hall, Princeton University, Princeton, NJ 08544}

\newcommand{\pitt}{Department of Physics and Astronomy, University of Pittsburgh, 
 Pittsburgh, PA 15260}
 
 \newcommand{\ppacc}{{Pittsburgh Particle Physics, Astrophysics, and Cosmology Center, University of Pittsburgh, Pittsburgh PA 15260}}
 
 \newcommand{\astrofisicacatolica}{Instituto de Astrof\'isica and Centro de Astro-Ingenier\'ia, Facultad de F\'isica, Pontificia Universidad Cat\'olica de Chile, Av. Vicu\~na Mackenna 4860, 7820436 Macul, Santiago, Chile}
 
 \newcommand{\oxford}{Sub-Department of Astrophysics, University of Oxford, Keble Road, Oxford, UK OX1 3RH}
 
\newcommand{\stonybrook}{Physics and Astronomy Department, Stony Brook University, Stony Brook, NY  11794}
\newcommand{\michigan}{Department of Physics, University of Michigan, Ann Arbor, USA 48103}
 \newcommand{\cornell}{Department of Physics, Cornell University, Ithaca, NY, USA 14853}
\newcommand{\cca}{Center for Computational Astrophysics, 162 5th Ave, New York, NY 10003}
\newcommand{\penn}{Department of Physics and Astronomy, University of
Pennsylvania, Philadelphia, PA 19104}
\newcommand{\NIST}{NIST Quantum Devices Group, 325 Broadway Mailcode
817.03, Boulder, CO, USA 80305}
%%%%%%%%%%%%%%%%%%%%%%%%%%%%%%%%%%%%%%%%%%%%%%%%%%%%%%%%%%%%%%%%%%%%%%%%%%%%%%%%%%%%%%%%%%

\author{Blake~D.~Sherwin}
\affiliation{Berkeley Center for Cosmological Physics, Lawrence Berkeley National Laboratory, Berkeley, CA,  94720}

\author{Alexander~van~Engelen}
\affiliation{\cita}

\author{Neelima~Sehgal}
\affiliation{\stonybrook}

\author{Mathew~Madhavacheril}
\affiliation{\princetonastro}
\affiliation{\stonybrook}

\author{Graeme~E.~Addison}
\affiliation{Dept. of Physics and Astronomy, The Johns Hopkins University, 3400 N. Charles St., Baltimore, MD,  21218-2686}

\author{Simone~Aiola}
\affiliation{\princetonphysics}
\affiliation{\pitt}
\affiliation{\ppacc}

\author{Rupert~Allison}
\affiliation{Institute of Astronomy, Cambridge, Madingley Road, Cambridge CB3 0HA, UK}

\author{Nicholas~Battaglia}
\affiliation{\princetonastro}

\author{Daniel~T.~Becker}
\affiliation{\NIST}

\author{James~A.~Beall}
\affiliation{\NIST}

\author{J.~Richard~Bond}
\affiliation{\cita}

\author{Erminia~Calabrese}
\affiliation{\oxford}
\author{Rahul~Datta}
\affiliation{\michigan}

\author{Mark~J.~Devlin}
\affiliation{\penn}

\author{Rolando D\"unner}
\affiliation{\astrofisicacatolica}
%\affiliation{Instituto de Astrof\'isica, Facultad de F\'isica, Pontificia
%Universidad Cat\'olica de Chile, Av.  Vicu\~na Mackenna 4860, 7820436 Macul, Santiago, Chile}

\author{Joanna~Dunkley}
\affiliation{\princetonphysics}
\affiliation{\princetonastro}
\affiliation{\oxford}

\author{Anna~E.~Fox}
\affiliation{\NIST}

\author{Patricio~Gallardo}
\affiliation{\cornell}

\author{Mark~Halpern}
\affiliation{Department of Physics and Astronomy, University of British Columbia, Vancouver, BC, Canada V6T 1Z4}

\author{Matthew Hasselfield}
\affiliation{Department of Astronomy and Astrophysics, The Pennsylvania State University, University Park, PA 16802}
\affiliation{Institute for Gravitation and the Cosmos, The Pennsylvania State University, University Park, PA 16802}

\author{Shawn~Henderson}
\affiliation{\cornell}

\author{J.~Colin~Hill}
\affiliation{Dept. of Astronomy, Pupin Hall, Columbia University, New York, NY 10027}

\author{Gene~C.~Hilton}
\affiliation{\NIST}

\author{Johannes~Hubmayr}
\affiliation{\NIST}

\author{John~P.~Hughes}
\affiliation{Department of Physics and Astronomy, Rutgers, The State University of New Jersey, Piscataway, NJ 08854-8019}

\author{Adam~D.~Hincks}
\affiliation{Department of Physics, University of Rome "La Sapienza", Piazzale Aldo Moro 5, I-00185 Rome, Italy}

\author{Ren\'ee~Hlozek}
\affiliation{\dunlap}

\author{Kevin~M.~Huffenberger}
\affiliation{Department of Physics, Florida State University, Tallahassee FL 32306}
\author{Brian~Koopman}
\affiliation{\cornell}
\author{Arthur~Kosowsky}
\affiliation{\pitt}
\affiliation{\ppacc}

\author{Thibaut~Louis}
\affiliation{UPMC Univ Paris 06, UMR7095, Institut d'Astrophysique de Paris, F-75014, Paris, France}

%\author{Jeff~McMahon}
%\affiliation{\michigan},

\author{Lo\"ic~Maurin}
\affiliation{\astrofisicacatolica}

\author{Jeff McMahon}
\affiliation{\michigan}

\author{Kavilan~Moodley}
\affiliation{\ukznastro}
\author{Sigurd~Naess}
\affiliation{\cca}
\affiliation{\oxford}

\author{Federico Nati}
\affiliation{\penn}

\author{Laura~Newburgh}
\affiliation{\dunlap}

\author{Michael~D.~Niemack}
\affiliation{\cornell}
\author{Lyman~A.~Page}
\affiliation{\princetonphysics}
%% \author{Bruce~Partridge}
%% \affiliation
%% \author{Jon~Sievers}
%% \affiliation

\author{Jonathan~Sievers}
\affiliation{Astrophysics and Cosmology Research Unit, School of
Chemistry and Physics, University of KwaZulu-Natal, Durban,
South Africa National Institute for Theoretical Physics,
KwaZulu-Natal, South Africa}

\author{David~N.~Spergel}
\affiliation{\princetonastro}
\affiliation{\cca}

\author{Suzanne~T.~Staggs}
\affiliation{\princetonphysics}

\author{Robert~J.~Thornton}
\affiliation{Department of Physics, West Chester University
of Pennsylvania, West Chester, PA 19383}
\affiliation{\penn}

\author{Jeff Van Lanen}
\affiliation{\NIST}

\author{Eve~Vavagiakis}
\affiliation{\cornell}
\author{Edward~J.~Wollack}
\affiliation{NASA/Goddard Space Flight Center, Greenbelt, MD 20771}

\definecolor{dark-gray}{gray}{0.3}
\newcommand{\tempcolor}[1]{\textcolor{black}{#1}}

\newcommand{\mnuConstraint}{\tempcolor{ 0.396 }}
\newcommand{\sigmaEightommConstraint}{\tempcolor{\ensuremath{0.643 \pm 0.054}}} %sigma8 * omega_M^(0.25)
\newcommand{\alensConstraint}{\tempcolor{\ensuremath{1.06 \pm 0.15~ (\mathrm{stat.}) \pm 0.06~ (\mathrm{sys.})}}}
\newcommand{\ommConstraint}{\tempcolor{\ensuremath{0.418 \pm 0.042}}}
\newcommand{\sigmaEightConstraint}{\tempcolor{\ensuremath 0.831 \pm 0.053}}
\newcommand{\overallpte}{\tempcolor{{\ensuremath 0.32}}}

\newcommand{\actpolDataFraction}{\tempcolor{12\%}}

\newcommand{\deepFiveNoise}{\tempcolor{12}}
\newcommand{\deepSixNoise}{\tempcolor{10.5}}
\newcommand{\actpolSignificance}{\tempcolor{7.1}}

\newcommand{\deepfivesixarea}{\tempcolor{626}}
\newcommand{\nmcsims}{\tempcolor{200}}
\newcommand{\nmeanfieldsims}{\tempcolor{400}}
\newcommand{\nbiassims}{\tempcolor{50}}
\newcommand{\morenbiassims}{\tempcolor{100}}
\newcommand{\pollensingsignificance}{\tempcolor{10}}

\newcommand{\alens}{\ensuremath{A_\mathrm{lens}}}

\newcommand{\curlpte}{\tempcolor{0.57}}%0.40

\newcommand{\caluncertainty}{\tempcolor{XX \pm YY}}
\newcommand{\dustscaling}{\tempcolor{0.04}}

\begin{abstract}
We report a measurement of the power spectrum of cosmic microwave background (CMB) lensing from two seasons of Atacama Cosmology Telescope Polarimeter (ACTPol) CMB data. The CMB lensing power spectrum is extracted from both temperature and polarization data using quadratic estimators.  We obtain results that are consistent with the expectation from the best-fit \planck\  $\Lambda$CDM model over a range of multipoles $L=80-2100$, with an amplitude of lensing $\alens=\alensConstraint$ relative to \planck. Our measurement of the CMB lensing power spectrum gives $\sigma_8\Omega_m^{0.25}=\sigmaEightommConstraint$; including baryon acoustic oscillation scale data, we constrain the amplitude of density fluctuations to be $\sigma_8 = \sigmaEightConstraint$. 
We also update constraints on the neutrino mass sum.
We verify our lensing measurement with a number of null tests and systematic checks, finding no evidence of significant systematic errors. This measurement relies on a small fraction of the ACTPol data already taken; more precise lensing results can therefore be expected from the full ACTPol dataset. 

\end{abstract}
\maketitle
\section{Introduction}

The large-scale structure of the Universe contains a wealth of information about the early universe, neutrinos, dark energy, and other physics that we are only beginning to extract. While measurements of large-scale structure using galaxies, quasars, Lyman-$\alpha$ absorbers, and other tracers continue to give great insight, these measurements are somewhat complicated by their reliance on biased probes of the mass distribution. In contrast, gravitational lensing directly probes all mass, including dark matter. 

The cosmic microwave background (CMB) radiation has unique advantages as a background light source for the study of gravitational lensing.  CMB photons originate from the last scattering surface at $z\simeq 1100$ and experience gravitational lensing deflections from large-scale structure along their paths to our telescopes. Hence, CMB lensing encodes information about nearly all the mass fluctuations in the Universe, with most of the signal arising between $z=0.5$ and $z=3$ \cite{blanchard87, bernardeau97, zaldarriaga98, lewischallinor}. The fact that much of the lensing signal originates from high redshifts and large scales means that the signal is simple to model, with most complications from non-linear evolution and baryonic physics negligible at current and near-future precision \cite{natarajan}. An additional simplifying feature is that the primordial CMB source is well understood, with a known redshift origin and simple statistical properties. Measurements of the CMB lensing signal therefore can serve as accurate probes of cosmology.

Given current measurement precision, the CMB lensing field can be modeled as Gaussian, so the power spectrum describes all its cosmological information; for future surveys, higher-order statistics may add information \citep{namikawa16,liu,boehm}. As the CMB lensing power spectrum probes the projected mass distribution, it is sensitive to both the growth of structure and the geometry of the Universe. Hence it is capable of constraining parameters such as neutrino mass, the amplitude of density fluctuations, curvature, and dark energy.

Measurements of the lensing power spectrum have only recently become possible with the advent of high-resolution, low-noise CMB telescopes such as the Atacama Cosmology Telescope (ACT) \cite{kosowskyatacama}, the South Pole Telescope (SPT) \cite{ruhlspt}, and the \planck\ satellite \cite{planckbluebook}. Following earlier cross-correlation results from WMAP \cite{smithlensing,hiratalensing}, the ACT team made the first measurement of the lensing power spectrum \cite{actlensing} and was able to confirm the existence of dark energy based on only CMB observations \cite{actparams}. The SPT collaboration was able to make a more sensitive measurement of temperature lensing \cite{engelenlensing}. The POLARBEAR collaboration made the first measurements of the lensing power spectrum using polarization data \cite{polarbearlensingA,polarbearlensingB}, following the first detection of polarization lensing in cross-correlation using SPTpol and {\it Herschel} \cite{hansonBmode}.  Subsequently, the SPTpol \cite{storylensing} and BICEP2/Keck \cite{biceplensing} teams presented measurements of polarization and temperature lensing power spectra with increased precision. The \planck\ team has made the current highest precision measurement of the lensing power spectrum: a 40$\sigma$ detection significance in their latest release \cite{plancklensing2013, plancklensing}.

While the \planck\  lensing power spectrum is generally in agreement with $\Lambda$CDM, the authors report some tension at small scales, with a null test failure at the $\sim 2.9\sigma$ level \cite{plancklensing}. In addition, several recent measurements using galaxy lensing and galaxy clusters have reported an amplitude of density fluctuations lower than that found with \planck\  lensing data, or \planck\  primary CMB data, at $2\sigma$ or higher significance, e.g. \cite{kidslensing}.  The main goals of this work are to present a new measurement of the lensing power spectrum, to independently constrain parameters such as the neutrino mass, and to introduce the ACTPol lensing pipeline. The possibility of testing both the \planck\  lensing results and any potential tensions between different measurements of the amplitude of structure provides additional motivation for our work.

This paper presents new measurements of the CMB lensing power spectrum using the first two seasons of ACTPol nighttime data and  the resulting constraints on cosmological parameters. The current measurement relies on only \actpolDataFraction\ of the usable ACTPol data already taken \cite{ACTPolSurveyStrategy}. Future measurements using the full ACTPol dataset will thus have higher precision, and our paper serves also as an exposition of the pipeline that we will use for this future work. Our analysis follows first-season ACTPol lensing results, which include a cross-correlation with maps of the cosmic infrared background fluctuations \cite[][]{engelencib};  a cross-correlation with radio sources to constrain their bias \cite{allison}; and a detection of lensing by dark matter halos by stacking on spectroscopic galaxies \cite{madhavacherillensing}. In section II, we describe the data and simulations we use in our analysis. In section III, we describe our pipeline for measuring the CMB lensing power. We present our results in section IV and verify our measurements with systematic estimates and null tests described in section V. We discuss the implications of our results for cosmological parameters in section VI and conclude in section VII.

\section{Data and Simulations}
ACT is a six-meter diameter CMB telescope operating in the Atacama Desert in Chile. The ACTPol receiver fitted to this telescope consists of three arrays of superconducting transition edge sensor bolometers, sensitive to both temperature and polarization; see \cite{actpolinstrument} for details on the instrument. ACTPol observed the sky at a frequency of 149 GHz in the first two years of the survey. The observations, data reduction and mapmaking are as described in the most recent ACTPol power spectrum analysis of Louis et al.~\cite{louis}, hereafter L16 (see also the previous analysis \cite{naess}). 

We use data taken in seasons 1 and 2 from three regions: D5 (57 $\mathrm{deg}^2$ at an effective white noise level of \deepFiveNoise~$\mu$K-arcmin) and D6 (71 $\mathrm{deg}^2$ at \deepSixNoise~$\mu$K-arcmin), both of which are contained within  a larger region, D56 (\deepfivesixarea{} ~$\mathrm{deg}^2$ at 17~$\mu$K-arcmin)\footnote{Although the maps are identical to those analyzed in \cite{louis}, the boundary of the analyzed regions differ slightly, leading to slightly different areas for each patch.}; see \cite{louis} for full noise spectra. These three regions are analyzed separately, because the significant variation in map depth would otherwise cause large statistical anisotropy that could be challenging to simulate and subtract accurately. Because the deep survey regions are located entirely within the wide survey footprint, the three different maps cannot be treated as statistically independent in our analysis.

Each of the fields is further processed to reduce the effect of resolved point sources and bright SZ clusters.  Our method for this follows the first-season analysis described in van Engelen et al.~\cite{engelencib}.  First, we template subtract the detected point sources to a flux limit of 5 mJy.  In the temperature maps, we additionally in-paint extended galaxy cluster candidates detected at greater than $5\sigma$ significance (numbering 98 in D56), together with a small number (14 from two map-based catalogs from D56) of irregular, residual point sources detected at greater than 5$\sigma$ significance.  In the polarization maps, we mask sources detected at 20$\sigma$ (290 from two combined D56 catalogs).  In both cases we perform the in-painting using constrained Gaussian realizations of CMB and noise~\cite{bucherlouis}; the mask radii for this in-paint procedure are 5' for the clusters and the polarized sources, and 15' for the irregular sources. We apodize our maps using a mask constructed from a product of the  weight map, smoothed with a Gaussian of width $l=1200$ in Fourier space, and a cosine-squared edge roll-off of total width 1.7$^o$, where the weights are proportional to the number of detector hits on each map pixel. All maps are deconvolved by the appropriate beams. The resulting polarization maps in Stokes parameters Q and U are then transformed to the $E-B$ basis using the pure-$EB$ method ~\cite{smithB}.  This method has been found to perform well for lensing reconstruction in~\cite{pearsonlens}.

Our simulations are generated as described in~\cite{actlensing} and~\cite{engelencib}.  To construct the signal component of our simulations, we create appropriately correlated, Gaussian-distributed $T$, $Q$, and $U$ primordial CMB maps using the best-fit parameters of \cite{calabreseparams}.  We then lens the maps with a Gaussian lensing potential using the algorithm described in \cite{louislensing}. We add Gaussian foreground power matching that from ACT observations as described in \cite{engelencib}. After convolving with the appropriate beam, each field is cut out of the larger CMB map; this ensures that the cut-out fields are correlated in the same way as our observed sky areas.

We construct the noise component of our simulations using the map hit-count and noise statistics from the data set as follows.  We make the map noise approximately isotropic by multiplying each pixel by the square root of the number of observations in that pixel; we then use 4 independent splits of the data to obtain a two-dimensional power spectral density, measuring it by subtracting the mean inter-split cross-spectrum from the mean auto-spectrum.  This power spectral density is then used to seed Gaussian random noise maps with the correct two-dimensional power spectral density.  The spatial inhomogeneity of noise levels over the map is modeled by dividing the simulations by the square root of the number of observations in each pixel. 

After adding the noise and signal components together, the simulations are apodized and transformed into the $E$ and $B$ polarization basis in exactly the same manner as the data.  We generate 400 simulations of each field using this method. The full simulation power spectra were found to match those of the data to within $\approx 5\%$ (the high-statistical-weight temperature map of D56 having the best match of $3\%$, and the low-weight D5/D6 polarization maps having the worst match to within $10\%$). We also generate 400 simulated maps with the same lensing potential realizations as the original simulation set, but with different background CMB and noise realizations, which we use to calculate higher-order lensing biases (as first implemented by \cite{namikawabias} and as described in the following section).

\section{Lensing Pipeline}
In this section, we describe our method to estimate the CMB lensing power spectrum. The methodology in our pipeline is overall similar to that presented in \cite{plancklensing, storylensing, engelencib}.

Since a fixed projected dark matter map introduces statistical anisotropy into the CMB by gravitational lensing, CMB lensing introduces correlations between formerly independent Fourier modes of the CMB temperature and polarization fields. Exploiting these lensing-induced correlations between pairs of modes, we can reconstruct the lensing potential with quadratic estimators in the CMB temperature and polarization fields $X = \{T, E, B\}$ \cite{huokamoto}:
\beq
\bar \phi^{XY}_\bL = R_\phi^{XY}(\bL) \int \frac{d^2\bl}{(2 \pi)^2}  X(\bl) Y(\bL-\bl) g_\phi^{XY}(\bl,\bL)
\label{eq:norm}
\eeq
where $g$ is a weighting function on the modes used in the quadratic estimators and $R_\phi$ is a normalizing function obtained analytically following \cite{huokamoto}.  The estimators we consider in our analysis are $XY = \{TT, TE, EE, EB\}$, because the $TB$ estimator has negligible signal-to-noise and the $BB$ correlation is higher order. The two CMB maps $X, Y$ we use in the estimators have been filtered to include only scales $1000<|\bl|<3000$. For a detailed discussion of this choice, focusing in particular on the minimum value of $|\bl|$ used, see Appendix~\ref{sec:appendix}. In addition, `stripes' of width $-90<|\bl_x|<90$ and $-50<|\bl_y|<50$ have been removed from the maps along Fourier axes corresponding to map declination and right ascension, respectively.

The function $g_\phi^{XY}(\bl,\bL)$ provides an optimal weighting given by the mean response of a pair of CMB fields $X(\bl) Y(\bL-\bl)$ to a lens $\phi_\bL$, divided by the variance of this pair of fields. The simplest example is given by the $TT$ estimator, for which
\beq 
g_\phi^{TT}(\bl,\bL) = \frac{C^{TT}_l {\bl \cdot \bL} + C^{TT}_{|\bL-\bl|} {\bL \cdot (\bL - \bl)} }{2(C^{TT}_l+N^{TT}_\bl)(C^{TT}_{|\bL-\bl|}+N^{TT}_{\bL-\bl})}
\eeq 
where $C^{TT}_l$ is the temperature power spectrum including the peak smearing from lensing and $N^{TT}_\bl$ is the temperature noise power spectral density. Analogous expressions for the other estimators can be found in \cite{huokamoto}, though we follow \cite{hanson2010} and replace unlensed with lensed spectra in filters to cancel higher-order biases. 

The normalization function $R_\phi^{XY}(\bL)$ divides out the weights $g$ to ensure an unbiased estimator. As a first approximation, it is calculated analytically as in \cite{huokamoto}. For example, for the $TT$ estimator, our first approximation to $R_\phi	^{TT}$ is 
\beq 
R_\phi^{TT}(\bL) \approx L^2 \left[\int \frac{d^2\bl}{(2 \pi)^2}  \frac{\left(g_\phi^{TT}(\bl,\bL)\right)^2}{(C^{TT}_l+N^{TT}_\bl)(C^{TT}_{|\bL-\bl|}+N^{TT}_{\bL-\bl})}\right]^{-1}.
\label{eqnorm2}
\vspace{0.3cm}
\eeq 
We apply a small correction to this function  when binning the estimator  in $L$-space (e.g., to account for windowing effects); this correction is obtained by requiring that the cross-correlation of the reconstructed lensing field with the input lensing field from simulations recovers the input lensing power spectrum of the simulations. In calculating this normalization correction, since two powers of the data mask enter into the quadratic lensing estimator, we apodize the input lensing potential simulations with the square of the data mask to mimic and absorb aliasing affecting the lensing reconstruction. For each estimator, the integrand of Eq.~\ref{eq:norm} can be written as sums of different convolutions of two Fourier space maps, so that the convolutions of Eq.~\ref{eq:norm} can be calculated using real space multiplications of different filtered fields \cite{huokamoto}. The use of inverse FFTs in evaluating the integrals of Eq.~\ref{eq:norm} (and similarly, Eq.~\ref{eqnorm2})  allows us to greatly speed up the lensing estimation and improve the scaling of computer time with map size. 
We assume the flat-sky approximation in our analysis, which is sufficiently accurate for the map sizes we use and the range of scales we seek to reconstruct \cite{actlensing, engelencib}. 

Even in the absence of lensing, anisotropic noise and window functions will produce spurious statistical anisotropy that affects the naive lensing estimator. An unbiased estimate of the potential can be recovered by subtracting this anisotropy signal, known as the mean field $\langle \bar \phi^{XY}_{\bL} \rangle$, that is induced by these types of non-lensing mode couplings. This mean field correction is calculated by averaging the reconstructions of the naive estimator from \nmeanfieldsims ~simulations, each with independent CMB and lensing potential realizations. In this average, only the spurious, non-lensing mode couplings remain. We then recover an unbiased estimate of the lensing potential after subtracting this mean field:
\beq
\hat \phi^{XY}_{\bL} =  \bar \phi^{XY}_{\bL} - \langle \bar \phi^{XY}_{\bL} \rangle.
\eeq
We use barred variables to indicate biased estimators. 

From these potential maps, we calculate the lensing power spectrum using the following naive estimator:
\beq
\bar C^\phi_{L_b} [XY,AB] \equiv \frac{1}{w_4} \sum_{b} \langle \hat \phi_\bL^{XY^*} \hat \phi_\bL^{AB} \rangle 
\label{binEquation}
\eeq
where $XY$ and $AB$ can be any of $TT, TE, EE, EB$, and the factor $w_4$ is calculated by taking each pixel value of the apodization mask to the fourth power and then averaging over pixels.  To maximize the signal to noise ratio, the  bandpowers are  binned using a  weight in the two-dimensional Fourier plane  $\bL$ given by the fiducial signal and noise spectra using $\left(C^\phi_L /( C^\phi_L + R_\phi(\bL))\right)^2$. Since each $\hat \phi^{XY}_{\bL}$ is a quadratic estimator in temperature and polarization, $\bar C^\phi_{L_b}$ is a four-point function in the CMB fields.

This naive lensing power spectrum estimate, Eq.~\ref{binEquation}, is biased, however, because a contribution to the lensing reconstruction power arises from both instrumental noise and the primary CMB. To obtain an unbiased estimator, this reconstruction ``noise bias'' must be subtracted. This bias can be understood if we consider averaging over the lensing field in addition to the background CMB; the measured power is comprised of a non-Gaussian (connected) part of the four-point function and a Gaussian (disconnected) part.  The former is the lensing power spectrum of interest, and the latter must be subtracted off.  We will refer to this bias term as the Gaussian bias (though it is often referred to as the ``$N_0$'' bias). 

In addition to the Gaussian bias, a bias must be subtracted that arises from additional connected contractions of two lensing potential fields in the measured four-point-function, contributing at first order in the lensing power, and known as the ``$N_1$'' bias. Furthermore, a small ``Monte Carlo'' ($\mathrm{MC}$) bias must be simulated and subtracted to absorb any additional non-idealities not captured by the map-level mean field subtraction, for instance due to masking correlations beyond the mean field or higher-order corrections (we choose to treat this correction as additive, which is sufficient for small corrections). 
Finally, we also subtract a small modeled foreground bias $\Delta C^{FG}_{L_b}$ ($3\%$ of the signal for temperature) from unresolved point sources and galaxy clusters, as detailed in Section~\ref{sec:syst}.  For each temperature and polarization combination, after subtracting off the Gaussian, $N_1$, $\mathrm{MC}$, and foreground biases, the final unbiased estimate of the lensing power spectrum that we use for our analysis is given by \cite{plancklensing2013, storylensing}:
\beqn
\hat C^\phi_{L_b} [XY,AB] &=&\bar C^\phi_{L_b} [XY,AB] \\\nonumber&-& \Delta C^{\mathrm{Gauss}}_{L_b} [XY,AB]\\\nonumber&-& \Delta C^{N1}_{L_b} [XY,AB] \\\nonumber&-&\Delta C^{\mathrm{MC}}_{L_b} [XY,AB]\\\nonumber&-&\Delta C^{\mathrm{FG}}_{L_b} [XY,AB]
\eeqn
where $\bar C^\phi_{L_b}$ is the biased estimate.

The biases we have described above are calculated as follows. The Gaussian ($N_0$) bias is calculated using the method described by \cite{namikawabias} from different pairings of data and simulation (superscript $S$) maps:
\beqn
\Delta C^{\mathrm{Gauss}}_{L_b} [XY,AB]&& = \nonumber \\ \langle \bar C^\phi_{L_b} [XY^S,AB^S] &+& \bar C^\phi_{L_b} [X^SY,AB^S]  \\\nonumber
+\bar C^\phi_{L_b} [X^SY,A^SB] &+& \bar C^\phi_{L_b} [XY^S,A^SB] \\-\bar C^\phi_{L_b} [X^SY^{S'},A^SB^{S'}] &-& \bar C^\phi_{L_b} [X^SY^{S'},A^{S'}B^S]\nonumber \rangle_{S,S'}.
\eeqn
This method is constructed to self-correct for small differences between the two-point functions of simulations and data. This is accomplished by using the two-point correlation functions of the data, rather than the simulations alone, to calculate the Gaussian bias; the first four terms each isolate a different two-point contraction of data maps. The robustness to incorrect simulations can be seen in detail by expanding the two-point correlation function of the data about the two-point function of the simulation, and noting that the difference in this expression cancels to first order, as demonstrated in BICEP/Keck 2016 \cite{biceplensing}.  The detailed form of the estimator can be obtained by deriving an optimal trispectrum estimator from an Edgeworth expansion of the CMB likelihood \cite{namikawabias}. Aside from providing robustness to simulations not matching real data, the use of a realization-dependent bias derived from data in the bias estimation  has additional advantages such as reducing the correlation of different lensing potential bandpowers \cite{hanson2010} as well as reducing the correlation of the lensing power spectra with the primary CMB spectra \cite{schmittfull,green,peloton}. 

To calculate the average over simulations we use \morenbiassims~different realizations of the simulation pairs $S, S'$ to obtain this bias for each real and simulated measurement. We verify that increasing the number of different realizations from \nbiassims~to \morenbiassims~did not substantially affect the results (with the amplitude only changing by $\sim1$\%). We therefore conclude that \nbiassims~realizations are sufficient for convergence, though we use \morenbiassims~in our data to be conservative.

\begin{figure}[t]
\includegraphics[width=\columnwidth]{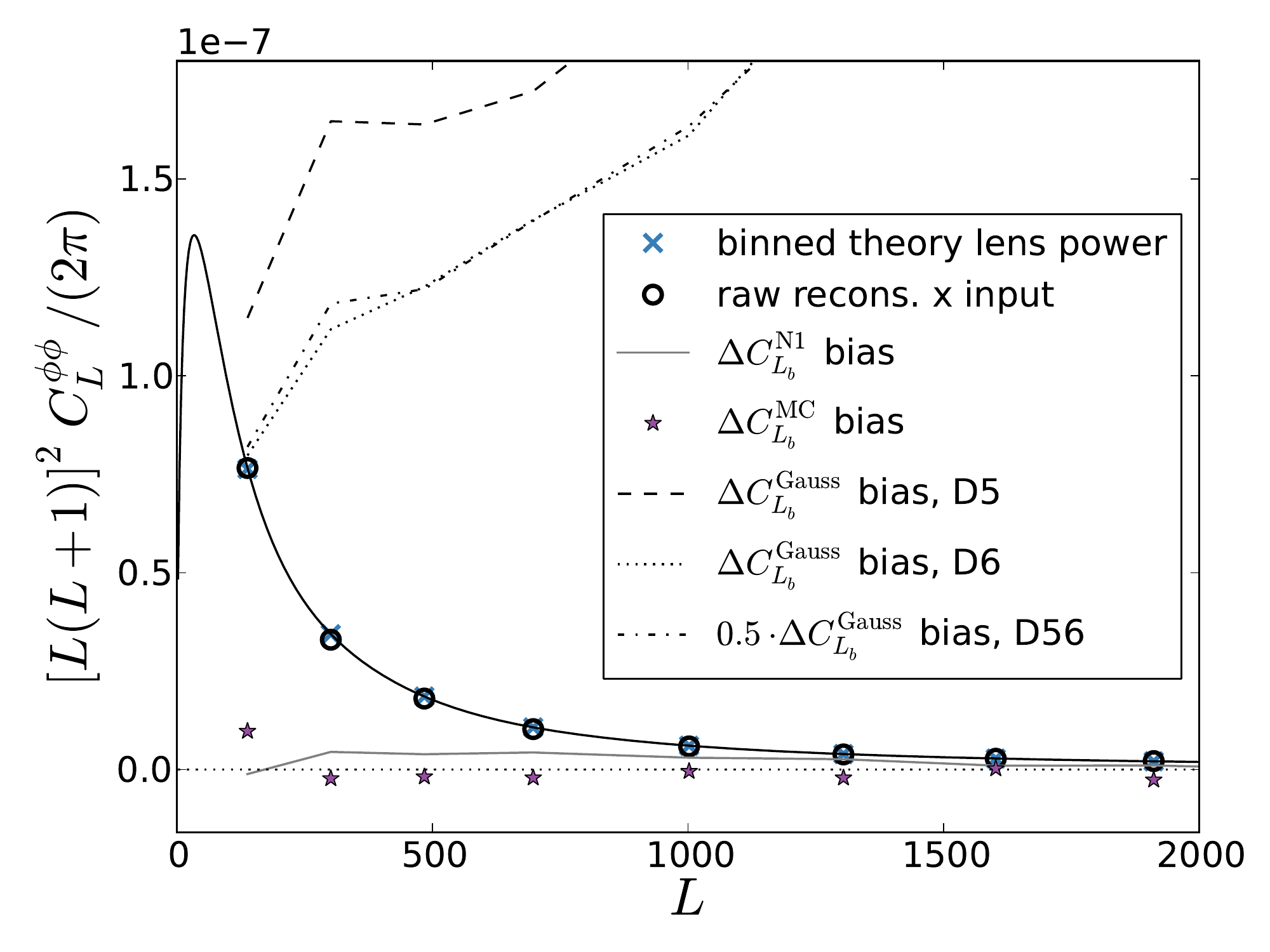}
\caption{Bias and signal terms from our pipeline simulations. The solid black circles show that the raw reconstructed maps correlate well with the input simulations (even without a simulation-based correction), closely matching the binned theory, denoted with light blue crosses. The dashed, dotted, and dot-dashed lines give the Gaussian ($N_0$) bias for the D5, D6, and D56 patches respectively. The $N_1$ bias, indicated with a solid gray line, has a magnitude and shape consistent with expectations. The MC bias, indicated with stars, is small as expected. All curves are co-adds over estimators and, where not indicated otherwise, over all patches, with the weights of Eq.~\ref{weightEqn}.}
\label{fig:verification}
\end{figure}

We obtain the $N_1$ bias by using simulations with different CMB realizations, but common lensing potential maps, following \cite{storylensing}:
\beqn
\Delta C^{\mathrm{N1}}_{L_b} [XY,AB]  &&= \nonumber \\ 
\langle \bar C^\phi_{L_b} [X^{S_\phi}Y^{S_\phi'},A^{S_\phi}B^{S_\phi'}]&+& \bar C^\phi_{L_b} [X^{S_\phi}Y^{S_\phi'},A^{S_\phi'}B^{S_\phi}] \\
\nonumber -\bar C^\phi_{L_b} [X^SY^{S'},A^SB^{S'}] &-& \bar C^\phi_{L_b} [X^SY^{S'},A^{S'}B^S]\nonumber  \rangle_{S,S', S_\phi,S_\phi'}.
\label{eq:n1}
\eeqn
Here $S_\phi$ and $S_\phi'$ indicate simulations with different CMB realizations but the same lensing potential realization, whereas $S$ and $S'$ have different CMB and lensing potential realizations.  Though subtracting the Gaussian and $N_1$ biases should result in a nearly unbiased lensing power spectrum, the Monte Carlo bias $C^{\mathrm{MC}}_b [XY,AB]$ is obtained by calculating any remaining residual from simulations. We find this MC bias to be significantly smaller than our one sigma error, so we use it in our pipeline although its inclusion does not substantially change our results.

Finally, we combine our estimate of the lensing potential power spectrum for all three patches $p$ and all ten estimators $\alpha=\{ TT,TT;~~TT,TE;~~ TT, EE \cdots\}$:
\beq
\hat C^\phi_{L_b} =  \sum_{\alpha,p} w_{\alpha,p,b} \hat C^{\phi,\alpha,{p}}_{L_b}
\label{weightEqn}
\eeq
Here the weights for  the  bandpowers of each estimator and patch are given by the inverse of the variance of the bandpowers, obtained from \nmcsims~simulations. While our  weights do not take into account correlation of the different estimators and maps,  our final coadded error calculation does, because it simulates the full measurement and coadding procedure. We calculate error bars and a full covariance matrix for the final bandpowers by repeating the complete coadd power spectrum estimation procedure on \nmcsims~simulations. We discuss our systematic error estimate in Section V of this paper.

A plot showing the relevant bias terms for our pipeline, along with additional information useful for verification, is given in Figure~\ref{fig:verification}. Even before correcting the normalization function $R$ with simulations, we note that the cross-correlation of the raw reconstructed lensing field with the input lensing field from simulations matches the input lensing power spectrum of the simulations to better than $5\%$. In addition, we find the N1 bias to have the expected form and the MC bias to be small, which gives us further confidence.

\begin{table}[t!]
\begin{threeparttable}
\caption{ACTPol two-season lensing power spectrum bandpowers and 1-$\sigma$ error bars.}
\centering
\input{latexTableActpol2016}
\label{tab:bandpowers}
\end{threeparttable}
\vspace{0.2cm}
\end{table}

\begin{figure*}[t!]
\includegraphics[width=1.5\columnwidth]{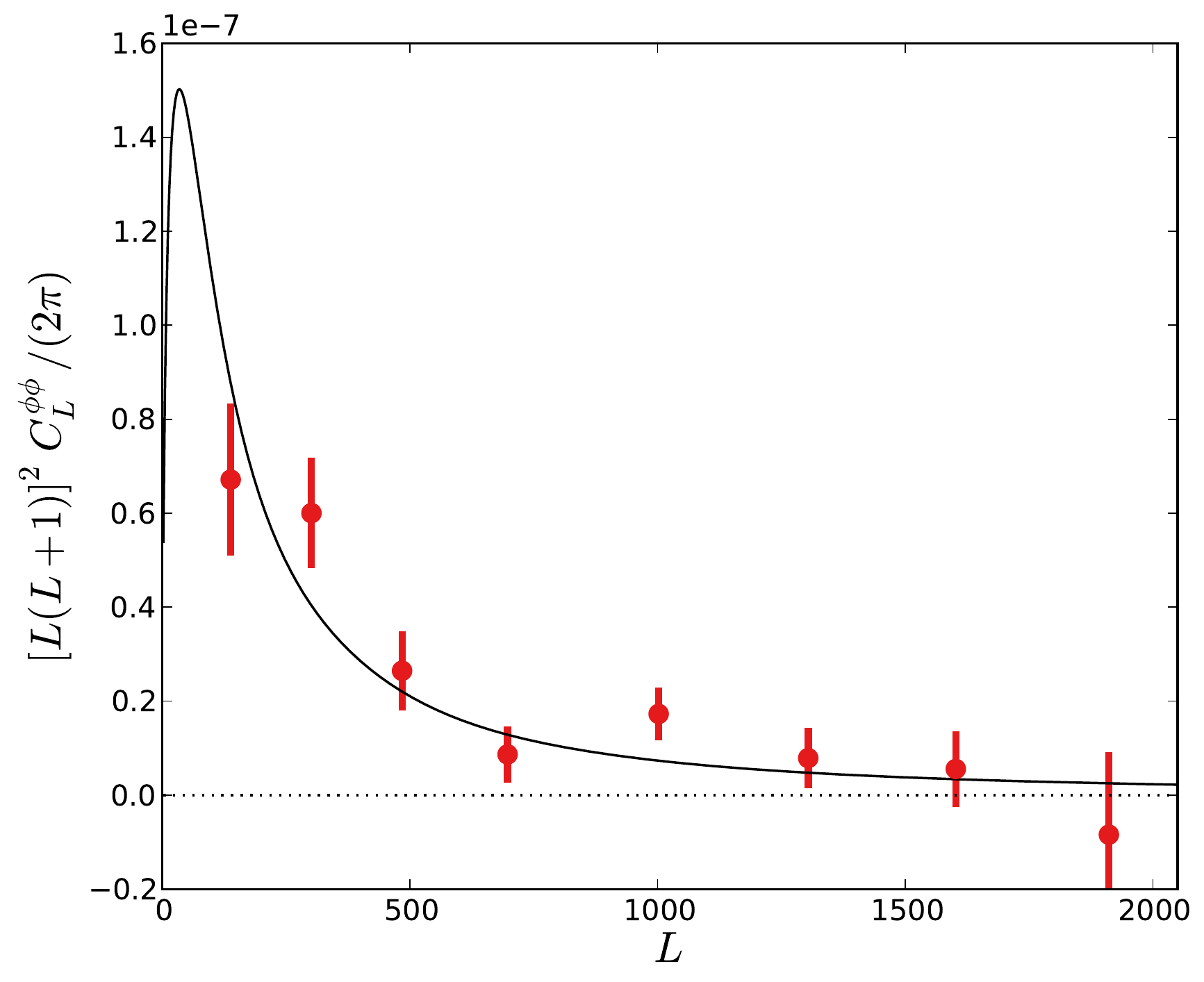}
\caption{Combined two-season ACTPol lensing power spectrum, coadded across all patches and estimators. The best-fit theory lensing power spectrum has an amplitude of $\alens=\alensConstraint$ relative to the \planck\ best-fit $\Lambda$CDM cosmology from the \planck\ temperature and polarization power spectra (which we define to have $\alens = 1$).  The ACTPol best-fit is indicated with a black solid line, and the error bars just include statistical uncertainty.  The $\chi^2$ to the best-fit, scaled \planck\  $\Lambda$CDM theory model has a probability to exceed (PTE) of \overallpte, suggesting a good fit to the standard $\Lambda$CDM cosmology.}
\label{fig:kappaAutoPlot}
\vspace{0.4cm}
\end{figure*}

\section{Lensing Power Spectrum}
In Figure~\ref{fig:kappaAutoPlot}, we show the final lensing power spectrum coadded over all estimators and patches, and in Table \ref{tab:bandpowers} we give the bandpower values and error bars. The amplitude of lensing power we obtain from the coadded result in Figure~\ref{fig:kappaAutoPlot}, scaled from the \planck\   TT,TE,EE+lowP+lensing $\Lambda$CDM model of \cite{plancklensing}, is \alens~of $\alensConstraint$. This represents a $7.1\sigma$ measurement of the amplitude of lensing. %% Lensing is thus detected at a combined significance of \actpolSignificance sigma in the ACTPol data. 
Calculating a $\chi^2$ to our best-fit model, we obtain a probability to exceed (PTE) the given $\chi^2$ of \overallpte, indicating a good fit to $\Lambda$CDM. This lensing amplitude is consistent with, and slightly higher than, that in the standard \planck\ cosmology.  

\begin{table}[h]
\begin{threeparttable}
\caption{Probability to exceed the given $\chi^2$ for the lensing signal, compared to the best-fit model}
\centering
\begin{tabular}{|c | c c c|} % centered columns (4 columns)   
\hline 
 Estimator & D56 lens & D5 lens & D6 lens \\
\hline
            $TT,TT$  & 0.26          & 0.14                     & 0.91 \\
	    $TE,TE$  & 0.004       	  & 0.74         	     & 0.94\\ 
	    $EE,EE$  & 0.69       	  & 0.70         	     & 0.51 \\
	    $EB,EB$  & 0.42       	  & 0.84         	     & 0.94 \\
	    $TT,TE$  & 0.86       	  & 0.34        	     & 0.92 \\
	    $TT,EE$  & 0.92       	  & 0.79         	     & 0.22 \\
	    $TT,EB$  & 0.21       	  & 0.13       	     & 0.73 \\
	    $TE,EE$  & 0.84       	  & 0.88         	     & 0.64 \\
	    $TE,EB$  & 0.25       	  & 0.29         	     & 0.89 \\
	    $EE,EB$  & 0.77       	  & 0.92         	     & 0.92 \\
\hline
\end{tabular}
\label{tab:kappapte}
\end{threeparttable}
\end{table}

\begin{figure*}[h!]
\includegraphics[width=2.1\columnwidth]{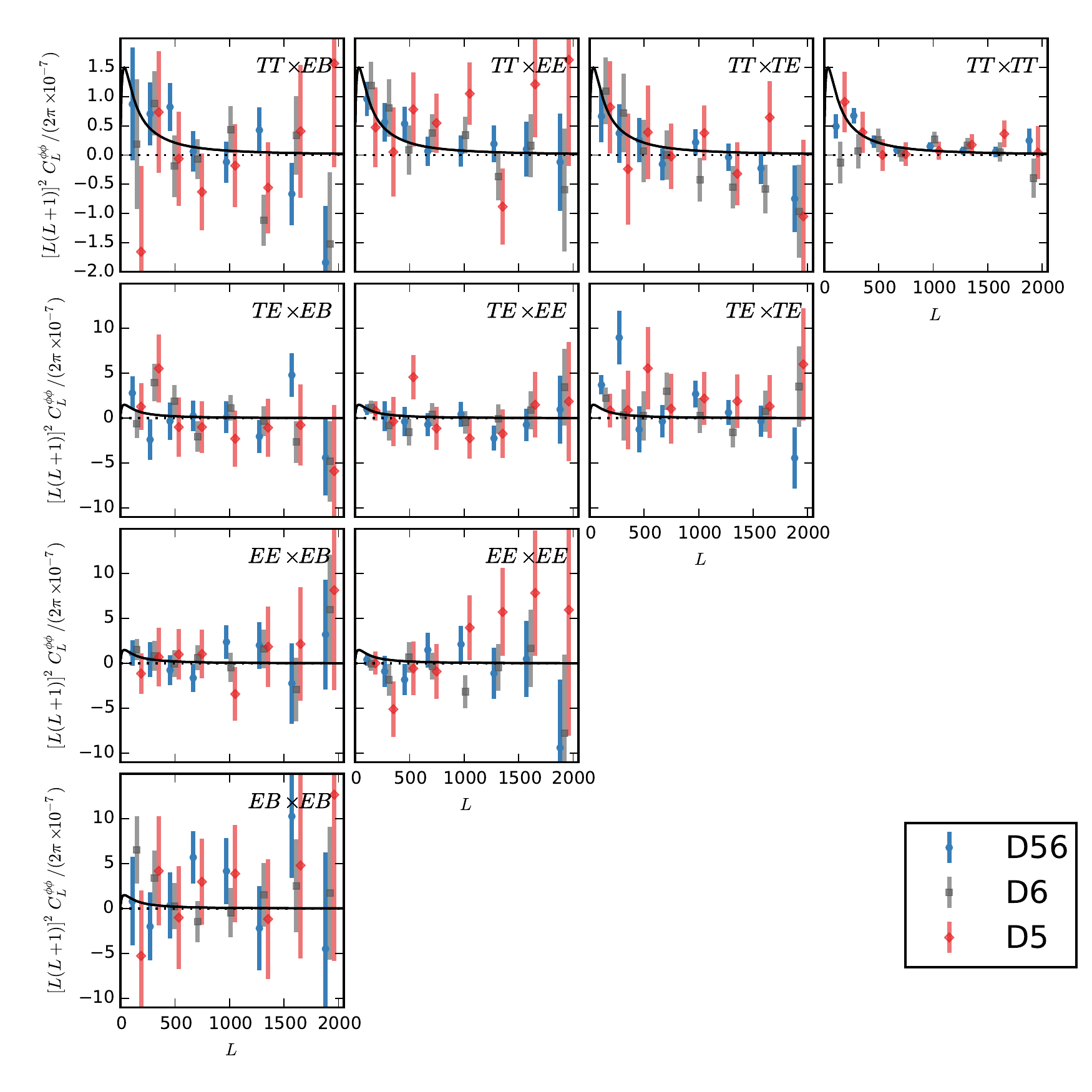}
  \caption{Lensing power spectrum reconstructions for all the different patches and estimators separately.  The blue points are from the D56 patch, the red from D5, and the grey from D6. We note that the fields overlap and are thus not independent.  The black curve is the best fit model.  The $TT,TT$ measurement (top right) can be seen to dominate the combined result, but the other combinations with temperature (top row) also have significant weight.  }
\label{fig:uberPlotAllkappa}
\end{figure*}

In Figure~\ref{fig:uberPlotAllkappa}, we show our results broken down by estimator and patch. From Figure~\ref{fig:uberPlotAllkappa}, it can be seen that most of the constraining power comes from the temperature data in the wider D56 map. 
In Table~\ref{tab:kappapte}, we list the individual PTEs for the lensing power from each estimator and patch. Though there is one entry, the $TE,TE$ estimator on D56, which has a low PTE of 0.44\%, we note that having a minimal PTE of this order is not unexpected, given that we calculate 30 signal PTEs and 30 null PTEs in this paper -- in fact, a mimimal PTE at or below this value occurs in 30\% of our simulated measurements. Excluding the D56 $TE,TE$ data shifts the best fit overall \alens~value downwards by only a small amount, approximately $0.25\sigma$ (to \alens = 1.02).

\section{Null Tests and Systematic Estimates}\label{sec:syst}

\begin{figure}[t!]
\includegraphics[width=\columnwidth]{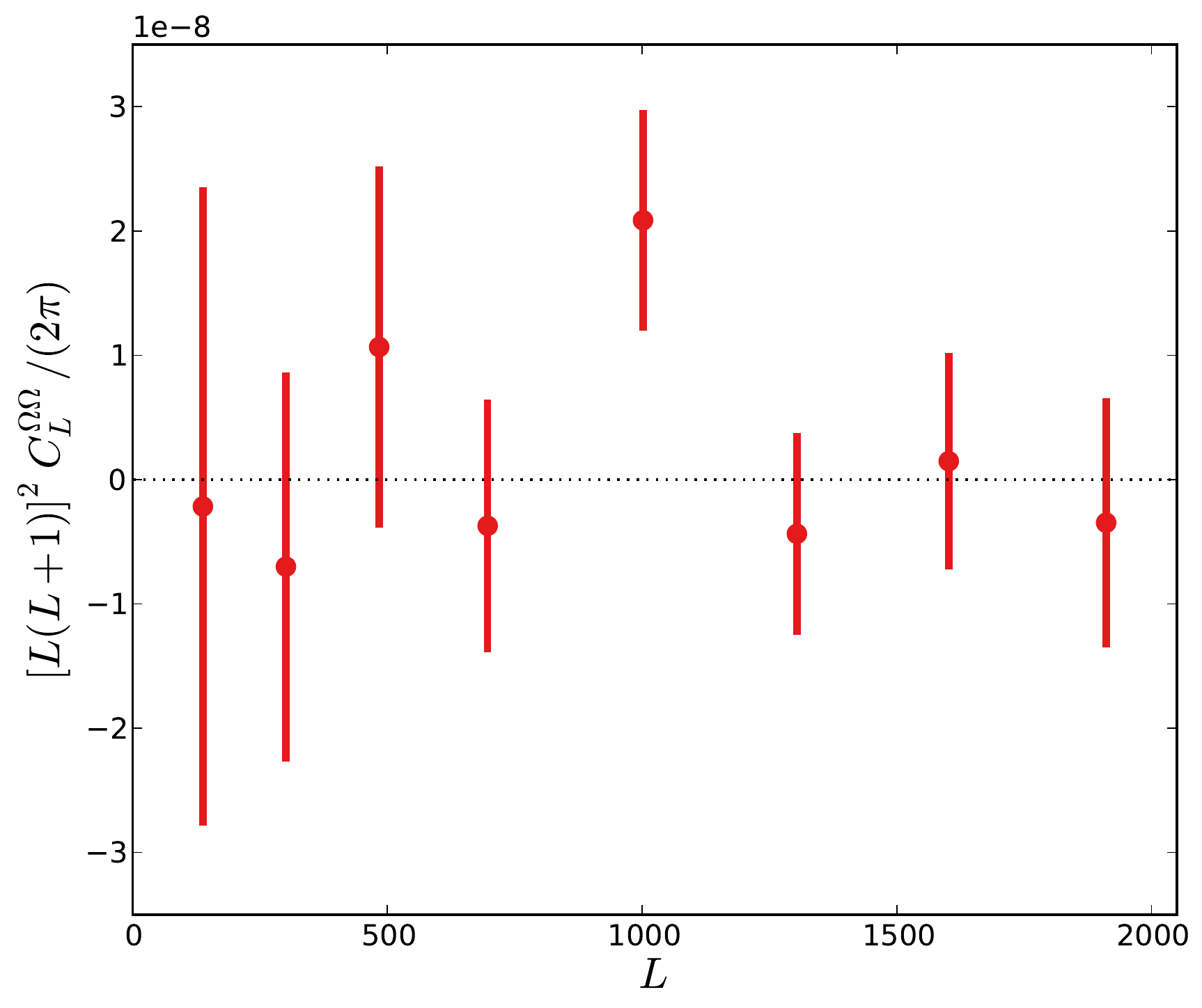}
\caption{Curl null test, coadded over all patches and estimators with the same weights as the lensing potential power spectrum. The PTE with respect to zero is \curlpte.}
\label{fig:curlCoadd}
\end{figure}

\begin{figure*}[h!]
\includegraphics[width=2.1\columnwidth]{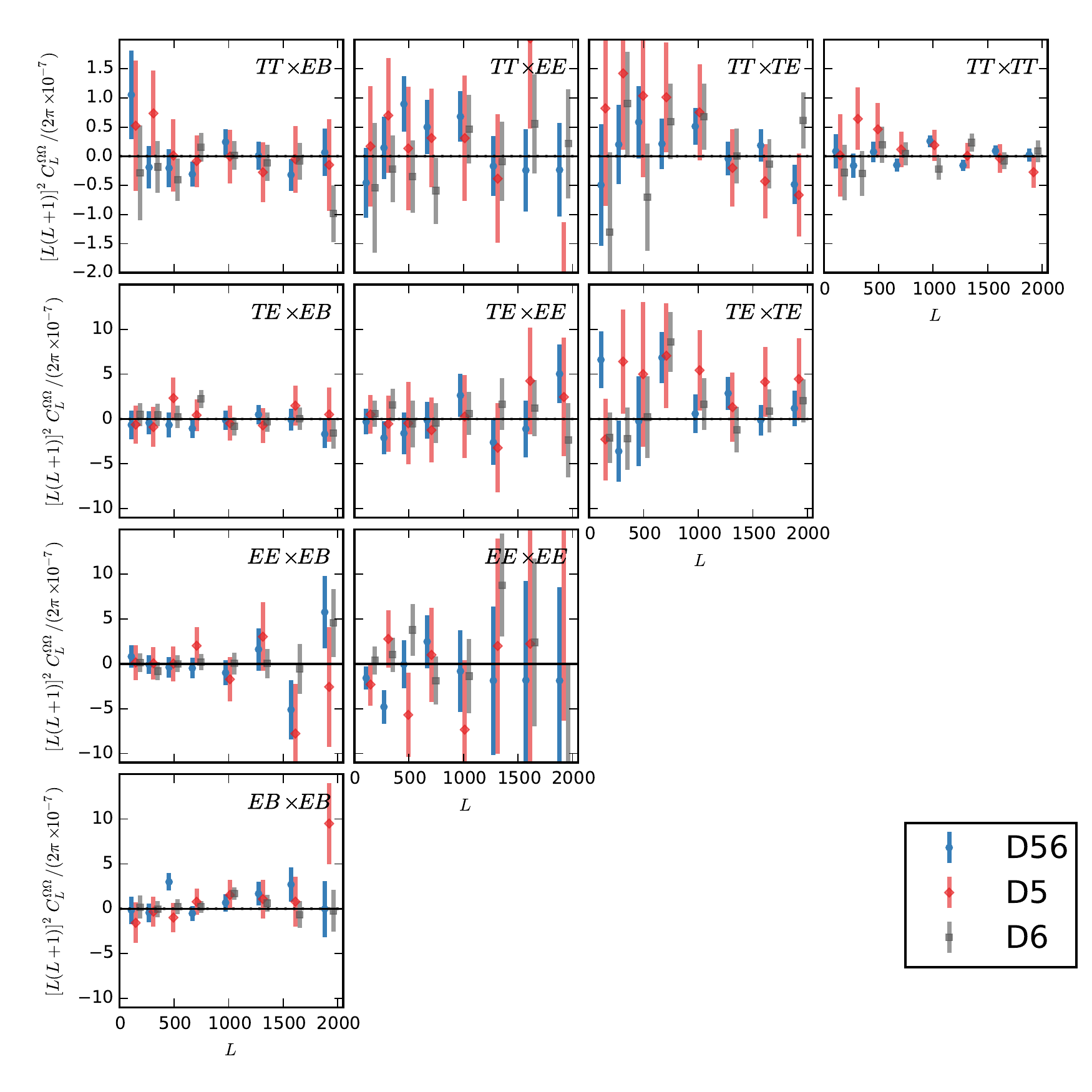}
\caption{Curl null test power spectra for all the different patches and estimators separately.  The blue points are from D56, the red points are from D5, and the grey points are from D6.} 
\label{fig:uberPlotAllomega}
\end{figure*}

We verify our results with null tests and targeted systematic checks. 
Neglecting very small corrections due to inadequacies of the lowest order Born approximation \cite{prattenlewis,marozzi}, the lensing deflection field is given by the gradient of the lensing potential from scalar density perturbations.  The deflection field hence is irrotational, with zero curl.  However, a systematic that mimics lensing need not necessarily obey this gradient-like symmetry, and could hence also induce a curl-like deflection. Estimating the curl-like component of the deflection field yields a  diagnostic for systematic errors that can mimic the lensing signal. We use a curl estimator given by
\beq
\Omega^{XY}_{\bL} = R_\Omega^{XY}(\bL) \int \frac{d^2l}{(2 \pi)^2}  X(\bl) Y(\bL-\bl) g_\Omega^{ XY}(\bl,\bL)
\eeq
where the filter $g_\Omega^{ XY}(\bl,\bL)$ differs from the usual lensing estimation filter by the replacement of a dot product in the numerator with the perpendicular component of a cross-product; the same modification occurs in the normalization function $R$. With this filter replacement, all the bias estimation steps are repeated in  the same way as for the lensing estimation. The results for this null test are shown in Figure~\ref{fig:uberPlotAllomega} for each estimator and patch separately, and in Figure~\ref{fig:curlCoadd} for the coadded result.   The curl PTEs for each estimator and patch with respect to zero are shown in Table~\ref{tab:curlnulls}, and are consistent with zero.  For the coadded curl, the PTE with respect to zero is~\curlpte, a good agreement with null.

\begin{figure}[t!]
\includegraphics[width=1.1\columnwidth]{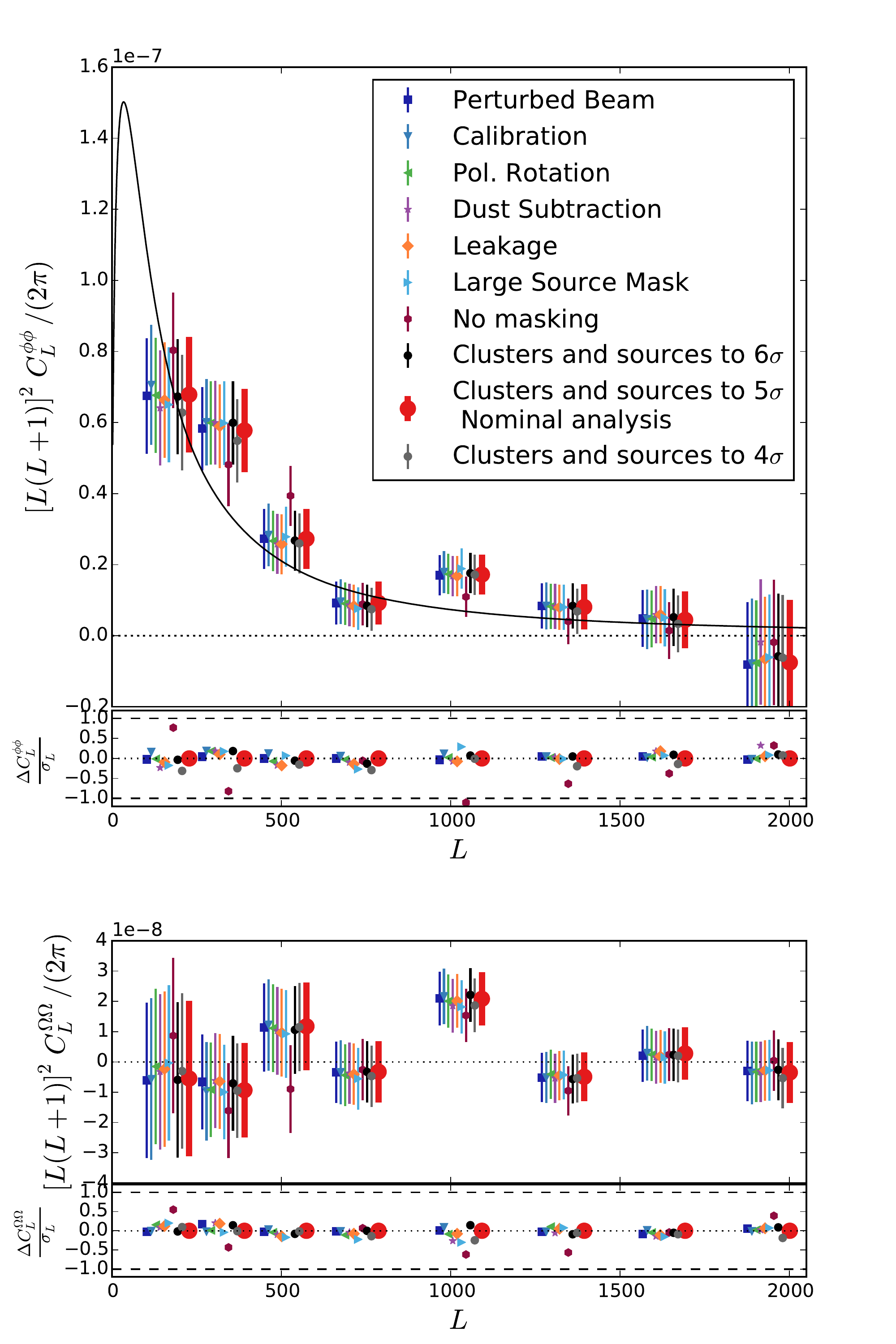}
\caption{Systematic test summary plot. The top panel shows the stability of our results to realistic levels of potential sources of systematic error, including instrumental errors in beams, calibration, polarization angle and temperature to polarization leakage, as well as astrophysical systematics such as galactic dust (see text for details). We also show the stability of our results when changing the masking threshold for SZ clusters and irregular point sources. The second panel shows the same data, but represented in terms of standard deviations.  We note the  stability of the results  to realistic levels of instrumental systematic effects and analysis choices (though it is apparent that some degree of source masking is required). The bottom two panels show the equivalent points for the curl null test. }
\label{fig:systematics}
\end{figure}

We also investigate the stability of our lensing power spectrum measurement to specific sources of systematic error. In  Figure~\ref{fig:systematics}, we show our measurement of the lensing power spectrum repeated with maps that have been perturbed by realistic levels of different sources of instrumental or astrophysical error. The sources of error we consider are described in the following paragraphs. For each potential systematic effect, we note the change in the best-fit lensing amplitude $\alens$, yielding an approximate estimate of its contribution to the total systematic error on our measurement.\\

\begin{table}[t!]
\begin{threeparttable}
\caption{Probability to exceed the given $\chi^2$ for the curl signal, compared to null}
\centering
\begin{tabular}{|c | c c c|} % centered columns (4 columns)   
\hline 
 Estimator & D56 curl & D5 curl & D6 curl \\
\hline
            $TT,TT$  & 0.10          & 0.89                     & 0.60 \\
	    $TE,TE$  & 0.05       	  & 0.66         	     & 0.29\\ 
	    $EE,EE$  & 0.43       	  & 0.74         	     & 0.58 \\
	    $EB,EB$  & 0.08       	  & 0.50         	     & 0.55 \\
	    $TT,TE$  & 0.49       	  & 0.75        	     & 0.45 \\
	    $TT,EE$  & 0.36       	  & 0.64         	     & 0.96 \\
	    $TT,EB$  & 0.53       	  & 0.99       	     & 0.54 \\
	    $TE,EE$  & 0.51       	  & 0.995         	     & 0.99 \\
	    $TE,EB$  & 0.90       	  & 0.98         	     & 0.43 \\
	    $EE,EB$  & 0.51       	  & 0.82         	     & 0.94 \\
\hline
\end{tabular}
\label{tab:curlnulls}
\end{threeparttable}
\vspace{0.2cm}
\end{table}

\noindent \emph{1. Beam uncertainty.}\\
We vary the beam within the uncertainties given in L16, coherently perturbing the beams in both temperature and polarization for all patches upwards by one standard deviation in order to obtain a conservative estimate. As shown in Figure~\ref{fig:systematics}, we find only small changes in the lensing bandpowers and a negligible overall shift of $\Delta \alens<0.01$. \\\\\\
\noindent \emph{2. Calibration uncertainty.}\\
We show the impact of CMB calibration uncertainty in Figure~\ref{fig:systematics}. As the limits quoted by L16 are $\approx$ 1\%, the bandpowers are perturbed by a factor $\approx 1.04$. The corresponding shift in the lensing amplitude is similarly $\Delta \alens  = 0.04$. We include this error as a contribution to the total systematic error on $\alens$.\\\\
\noindent \emph{3. Polarization angle uncertainty.}\\
We model a global polarization angle offset within the stated limits of L16  by adding 1\% of the $Q$ maps to $U$ and subtracting 1\% of the $U$ maps from $Q$. This results in small shifts to lensing bandpowers and a change in the overall amplitude of $\Delta \alens = 0.01$. We again include this value in our total systematic error budget. \\\\
\noindent \emph{4. Temperature-to-polarization leakage.}\\
We model instrumental temperature-to-polarization leakage by adding 1\% of the temperature map to the $E$-mode map (as a leakage of this form and magnitude was found to be present in initial versions of the ACTPol CMB maps, though it was fixed by better beam characterization, as described in L16).  We again find only small shifts to bandpowers and a change in the amplitude of lensing of $\Delta \alens = -0.02$, which we include in our total systematic error calculation.\\\\
\noindent \emph{5. Galactic dust.}\\
We calculate an upper bound on the impact of galactic dust by subtracting the \planck\  353 GHz maps below $\ell<2000$ from our CMB temperature maps (at higher $\ell$, CIB and instrumental noise become large and dominant). Prior to subtraction, we rescale the 353 GHz maps to serve as dust maps at 149 GHz by dividing by $\approx 20$ (see \cite{plancklensing}). We obtain a shift of $\Delta \alens = -0.03$, with only small changes in the lensing bandpowers. Though this value represents in some sense an upper bound (since a small fraction of the large-scale CMB is also removed), we include this value in our systematic error budget. Comparable small bounds were found in \cite{engelenlensing,plancklensing}. We note that the impact of polarized dust is expected to be very small, given that we only use information at $\ell>1000$ and given that most of our statistical weight is in the temperature estimator. Furthermore, we note that the curl null should be sensitive to an unexpectedly large dust bias \cite{storylensing}.\\\\
\noindent \emph{6. Source and cluster mask level and mask size.}\\
Steps have been taken in this analysis to mitigate the impact of astrophysical contaminants, such as observing in low-dust regions, masking and in-painting SZ clusters, and template-subtracting bright star-forming and radio galaxies. However, we also test for any effect on our results from residual astrophysical foregrounds. In Figure~\ref{fig:systematics}, we show the result of changing the number of masked clusters and residual sources, with mask thresholds corresponding to objects detected at $6\sigma$, $5\sigma$, and $4\sigma$ using a matched filter.  Our main result masks out SZ clusters and residual sources above $5\sigma$. The variation in bandpowers and in the amplitude of lensing is much less than the statistical error for all masking choices, with a root-mean-squared change of  $\Delta \alens = 0.03$ from the baseline result. 
We further test the stability of our results by doubling the size of the in-painting mask around each object. We find only small changes to bandpowers and an overall shift of  $\Delta \alens= -0.01$.  Finally, we display in Figure~\ref{fig:systematics} the lensing bandpowers when no masking of clusters and residual sources is performed. As expected, omitting the masking procedure entirely causes substantial shifts in the bandpowers. 
However, as shown in Figure~\ref{fig:systematics} our results are insensitive to the details of the masking procedure, which gives confidence in their robustness.\\

\begin{table}[t!]
\begin{threeparttable}
\caption{Systematic error budget. We list the different sources of systematic error investigated, along with an approximate (often conservative) estimate of their impact on the amplitude of lensing, $\Delta \alens$. Adding the different errors in quadrature, we obtain an estimate for our total systematic error, $\Delta \alens (\mathrm{sys.})$.}
\centering
\begin{tabular}{|c | c |} % centered columns (4 columns)   
\hline
Type of Systematic & Systematic Error, $\Delta \alens$\\
\hline
         Beams  &  $<$0.01         \\
	    Calibration  & 0.04       	  \\ 
	    Polarization Angle  & 0.01       	  \\ 
	    Temperature-Polarization Leakage  & 0.02       	  \\ 
	    Galactic Dust  & $<$0.03       	  \\ 
        Astrophysical (Clusters/Sources)& 0.03\\
\hline
Total Systematic Error  & 0.06         \\
\hline
\end{tabular}
\label{tab:systematics}
\end{threeparttable}
\end{table}

\noindent \emph{7. Unresolved astrophysical foregrounds}\\
Even with aggressive masking, some residual effective lensing signal will remain from the trispectra associated with extragalactic objects just below the cut threshold in the temperature maps \cite{osbornebias}.  These biases, arising from galaxy clusters, the cosmic infrared background, and radio sources, were estimated in \cite{engelenbias}, based partly on the simulations from \cite{sehgalsims}. We find that for our current masking levels and maximum multipole used in the reconstructions, the biases expected are roughly 3\% of the signal for the $TT,TT$ estimator. We use the relevant curves from \cite{engelenbias} as our foreground bias $\Delta C_{L_b}^{FG}$ that we subtract when deriving our final lensing power spectrum. For polarization, the foregrounds are expected to be much less of a concern -- SZ clusters produce only an extremely small polarized signal, and the polarized CIB and point source levels are also very small (e.g., L16). We thus neglect unresolved foreground biases in estimators using only polarization. For lensing power spectrum estimators involving one $TT$-estimator half (e.g, $TT,EB$), we assume a bias $\Delta C_{L_b}^{FG}$ given by one half of the $TT,TT$ bias (which is justified by a dominant contribution to the bias arising from the lensing-source-source bispectrum \cite{engelenbias}). If we turn off subtraction of all the bias, the amplitude of lensing shifts by $\Delta \alens = -0.02$. 

What contribution from astrophysical foregrounds such as galaxy clusters, CIB, and radio sources should we assign to our overall systematic error budget?  We note that there is of order $50\%$ theoretical uncertainty on the simulation-derived estimate \cite{engelenbias}, implying an error $\Delta \alens \approx  0.01$. To be conservative, we add (in quadrature) to this the dispersion found for different source and cluster masking levels, giving a total error of $\Delta \alens = 0.03$ for the astrophysical uncertainty in our measurement.\\

\noindent \emph{8. Noise tests}\\
Finally, we test our modeling of the noise. By differencing two splits of our data with equal weight and thus cancelling the signal, we obtain maps of the noise in our data. We add these maps to simulations of the lensed CMB signal and measure the lensing power spectrum of the resulting maps with our pipeline. The recovered lensing power spectrum is found to be a good fit to the input simulation power spectrum, with a PTE of 76\%. We repeat this analysis with a new realization of the background CMB signal, obtaining a PTE of 6\%. From different splits of our data, we also obtain a new, uncorrelated noise (and signal) map, which gives a PTE of 75\% in our test. For all three cases, the 30 individual PTEs for each patch and estimator combination appear nominal. We therefore find no significant evidence for systematics from noise modeling in our analysis.\\

In Figure~\ref{fig:systematics}, we also show the changes to the curl null test in response to the enumerated systematic effects. As none of the systematics or analysis choices we investigate (aside from not masking any sources at all) causes significant changes to the curl points, we conclude that the systematic effects investigated are not responsible for any features in the curl power spectrum. %We also conclude that the curl null test may not be an exhaustive probe of all plausible instrumental systematics.

We summarize the different sources of systematic error investigated, along with an approximate (often conservative) estimate of their impact on the amplitude of lensing, $\Delta \alens$, in Table \ref{tab:systematics}. By adding all these sources of error in quadrature, we obtain an estimate for the total systematic error on our measurement of the lensing power spectrum amplitude of $\Delta \alens (\mathrm{sys.})=0.06$.

For the systematic tests enumerated above, we note that the overall systematic error contribution is subdominant to the statistical error. In all tests, we do not see significant changes to our baseline results. Indeed, nearly all our estimates of systematics are conservative upper limits; there is no significant evidence for systematic contamination to our lensing measurement at the current level of precision from either astrophysical or instrumental effects.

\section{Cosmological Parameters}\label{sec:params}

\newcommand{\bitheta}{\boldsymbol{\theta}}
\newcommand{\be}{\begin{equation}}           
\newcommand{\ee}{\end{equation}}

In this section, we present cosmological constraints on the linear-theory matter fluctuation amplitude $\sigma_8$, the matter density $\Omega_m$, and the sum of the neutrino masses $\Sigma m_{\nu}$ from the ACTPol lensing power spectrum. We obtain these constraints from the coadded lensing power spectrum shown in Figure~\ref{fig:kappaAutoPlot}.

We model the ACTPol lensing likelihood by assuming Gaussian uncertainties on the correlated, binned 
coadded spectrum, $\hat C^\phi_{L_b}$, so that the log-likelihood is given by,

\be
-2\mathrm{ln}\mathcal{L} = \sum_{bb'}\left[\hat C^\phi_{L_b} - C_{L_b}^{\phi,\mathrm{th}}(\bitheta)\right]{\mathbb{C}}^{-1}_{bb'}\left[\hat C^\phi_{L_{b'}} - C_{L_{b'}}^{\phi,\mathrm{th}}(\bitheta)\right].
\ee
The Gaussian approximation is justified by the large number of effective independent modes in our bandpowers. We have checked that a correction due to having a finite number of simulations, based on \cite{hartlap}, yields only a 2-3$\%$ effect on our final bandpower errors. The covariance matrix $\mathbb{C}^{-1}$ for the binned spectrum is calculated using Monte-Carlo simulations as described in Section III. 
Since the normalization $R_\phi(L)$ in Eq.~\ref{eq:norm} and the $\Delta C^{\mathrm{N1}}_{L_b}$ bias correction in Eq.~\ref{eq:n1} assume a fiducial cosmology $\bitheta_0$, we calculate the expected spectrum, $C_{L}^{\phi,\mathrm{th}}(\bitheta)$, at the point $\bitheta$ in cosmological parameter space and correct it to reflect the $R_\phi(L)$ and $C^{\mathrm{N1}}_{L_b}$ we used for the data.  
%has a non-trivial dependence on the underlying theory power spectrum $C_{L}^{\phi}(\bitheta)$ because the normalization $R_\phi(L)$ in Eq~\ref{eq:norm} and the $\Delta C^{\mathrm{N1}}_{L_b}$ bias correction in Eq~\ref{eq:n1} assume a fiducial cosmology $\bitheta_0$. 
Since calculating the exact correction for each point in parameter space is prohibitively slow, we follow the approach in \cite{plancklensing} and exploit the near-linear dependence of the expected power spectrum, due to shifts in $R_\phi(L)$ and $\Delta C^{\mathrm{N1}}_{L_b}$, when expanding around the fiducial cosmology (see in particular Eq C.5 in \cite{plancklensing}). However, we neglect the contribution to the correction from the dependence of $\Delta C^{\mathrm{N1}}_{L_b}$ on the CMB primordial power spectra as these spectra are strongly constrained by the addition of CMB power spectrum information. In addition, the dependence of $\Delta C^{\mathrm{N1}}_{L_b}$ on the lensing power spectrum is assumed to be dominated 
by an overall scaling of the amplitude of the fiducial lensing power spectrum rather than on scaling each $L_b$-mode separately; this is a very good approximation for the parameters we consider, which effectively only smoothly rescale the lensing power spectrum. For any pair of estimators $XY, AB$ used for the power spectrum we therefore have
\beqn
C_{L,XYAB}^{\phi,\mathrm{th}}&=&C_{L}^{\phi}\\\nonumber&+&\frac{d\mathrm{ln}R_\phi^{XY} R_\phi^{AB}(L)}{dC^j_{\ell}}\Big|_{\bitheta_0}\left(C^j_{\ell}(\bitheta)-C^j_{\ell}(\bitheta_0)\right)C_{L}^{\phi}(\bitheta_0) \\\nonumber
& +& \Delta C^{\mathrm{N1}}_{L, XYAB}(\bitheta_0)\left(\frac{\langle C_{L}^{\phi}(\bitheta)\rangle}{\langle C_{L}^{\phi}(\bitheta_0)\rangle}-1\right).
\label{eq:corrections}
\eeqn
where $C_{L}^{\phi}$ is the theory power spectrum for  the given parameters, and where we estimate the mean amplitude of lensing $\langle C_{L}^{\phi}(\bitheta)\rangle$ by averaging $L$ times the lensing convergence power ($\sim L^5 C^{\phi}_{L}$) from $L=0-2000$.

The final theory lensing spectrum that is compared against the measured coadded lensing spectrum is the linear combination of the above spectra over all $XY, AB$ pairs, weighted and binned in the same way as the measured coadded lensing spectrum (Eq.~\ref{binEquation}).

\begin{table}[t]
\begin{threeparttable}
\caption{Priors used in the cosmological analysis when including and not including primary CMB temperature fluctuations}
\centering
\begin{tabular}{|c | c | c|} % centered columns (4 columns)   
\hline 
 Parameter & Without CMB $TT$ & With CMB $TT$  \\
\hline
\hline
            $\ln 10^{10}A_s$  & $[2,4]$ & $[2,4]$  \\
            $H_0$  & $[40,100]$ & $[40,100]$  \\
	    $n_s$  & $0.96 \pm 0.02$ & $[0.8,1.2]$     \\ 
	    $\Omega_bh^2$  & $ 0.0223 \pm 0.0009$ & $[0.005,0.1]$\\
	    $\Omega_ch^2$  & $[0.005,0.99]$ & $[0.005,0.99]$ \\
	    $\tau$  & $0.058 \pm 0.012$ & $0.058 \pm 0.012$ \\
	    $\sum m_{\nu}$ (eV)  & $0.06$ & $[0,10]$ \\
	    
	    \hline
\end{tabular}
\label{tab:priors}
\vspace{0.5cm}
\end{threeparttable}
\end{table}

We calculate theory power spectra using the Boltzmann code CAMB (using Halofit to model the effects of non-linear structure formation \cite{halofit2003, halofit2012}) and use the MCMC code CosmoMC \cite{cosmomc1,cosmomc2} to obtain parameter constraints.  We consider the basic six $\Lambda$CDM parameters - cold dark matter and baryon densities, $\Omega_c h^2$ and $\Omega_b h^2$, the optical depth to reionization, $\tau$, the Hubble constant, $H_0$, and the amplitude and scalar spectral index of primordial fluctuations, $A_s$ and $n_s$ - and a single family of massive neutrinos with total mass $\Sigma m_{\nu}$. These parameters are varied with priors as summarized in Table~\ref{tab:priors} and consistently with the Planck lensing analysis \cite{plancklensing}. We, however, update the $\tau$ estimate following more recent Planck data \cite{newplanckTau}. The prior on $\Omega_bh^2$ comes from big bang nucleosynthesis in combination with quasar absorption line observations \cite{PettiniCooke}, and the prior on $n_s$ is centered on {\it \planck\ } measurements of the CMB power spectra but with a relatively broad width \cite{planckParams2015}. 

As explained in detail in \cite{plancklensing}, the parameter combination that lensing measures best is $\sigma_8\Omega_m^{0.25}$. From ACTPol lensing alone, we obtain a constraint in the $\sigma_8$-$\Omega_m$ plane of 
\begin{align}
\sigma_8\Omega_m^{0.25}= \sigmaEightommConstraint \quad & (\text{ACTPol lens only, 68\%})
\label{eq:s8om}
\end{align}
%\ee
This is consistent with the \planck~lensing-only constraint of $\sigma_8\Omega_m^{0.25}= 0.591 \pm 0.021$~\cite{plancklensing}. Though the Planck lensing power spectrum measurement itself is much more precise than our measurement, the constraints on $\sigma_8\Omega_m^{0.25}$ are more comparable, because Planck's constraint on the $\sigma_8\Omega_m^{0.25}$ combination is degraded by marginalizing over $\Omega_m h^2$ and other parameters.

\begin{figure}[t!]
\includegraphics[width=\columnwidth]{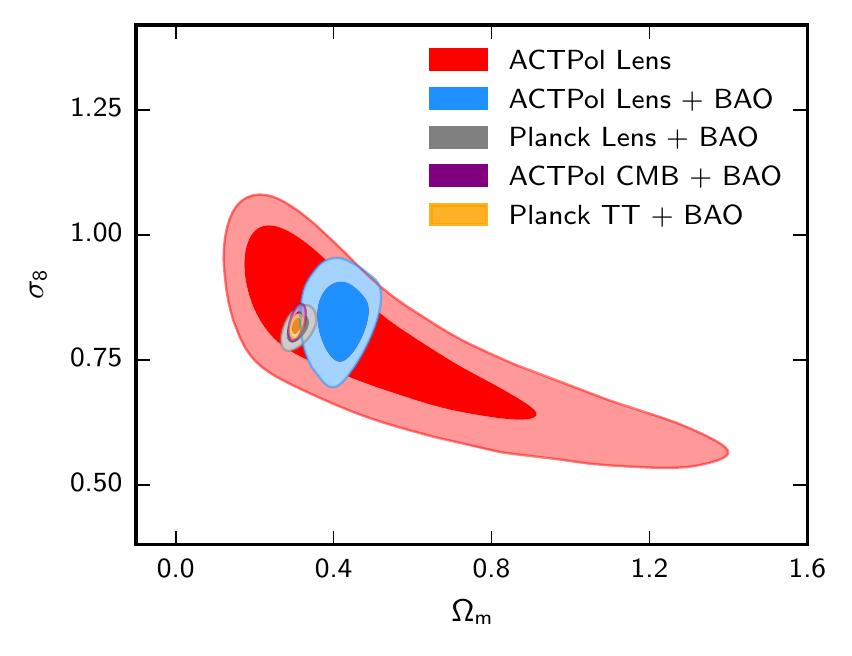}
\caption{Constraints at 68\%~CL on $\sigma_8$ versus $\Omega_m$ from ACTPol lensing data alone and in combination with BAO data.    
The red contours are from ACTPol lensing data (i.e. four-point correlation function information) alone when fixing the cosmology used in the lensing power spectrum normalization correction to the best-fit from \planck\  primary CMB data; this effectively restricts the background source plane to be consistent with primary CMB data.  The blue contour shows the result when BAO data is added (see text for details of the dataset). \planck\  four-point lensing data plus BAO is indicated by the grey contour.  For comparison with high redshift probes of the amplitude of structure, which do not necessarily probe the same physics in extensions to $\Lambda$CDM, we also show the contours from the CMB power spectra, for both ACTPol and \planck\  data plus BAO, as purple and yellow contours, respectively. These CMB power spectrum contours, which mainly probe the primary CMB, are consistent with our lower redshift lensing measurements.}
\label{fig:ACTOnly}
\end{figure}

Combining the ACTPol lensing likelihood with a BAO likelihood, which includes 6DF~\cite{sixdf}, SDSS MGS~\cite{mgs}, and BOSS DR12 CMASS and LOWZ data-sets~\cite{bossdr12}, we break the $\sigma_8$ - $\Omega_m$ degeneracy and obtain the following individual marginalized constraints,

%\be
\begin{align}
\sigma_8 = \sigmaEightConstraint \quad &(\text{ACTPol lens+BAO, 68\%}) 
\label{eq:s8}
\end{align}
%\ee

%\be
\begin{align}
\Omega_m= \ommConstraint \quad &(\text{ACTPol lens+BAO, 68\%}). 
\label{eq:om}
\end{align}
%\ee
We note that the constraints given in Eqs.~\ref{eq:s8om}--\ref{eq:om} are obtained while fixing the cosmology $\bitheta$ in the $R_\phi(L)$ correction given in Eq.~\ref{eq:corrections} to the \planck\  best-fit model from the \planck\  primary CMB data alone, just as done for the \planck\ lensing-only constraints obtained in \cite{plancklensing}. This restricts the statistics of the CMB background source light, giving a weaker constraint than fully adding the \planck\  primary CMB data to the ACTPol and BAO datasets.  If we allow the cosmology in the $R_\phi(L)$ correction to vary when only ACTPol lensing data are used, then the parameter chains explore regions of parameter space that are largely inconsistent with known measurements of the primary CMB, due to a degeneracy of the amplitude of the lensing signal with the CMB power spectra (in our case, primarily with an integral scaling as $(C_{\ell}^{TT})^2$).

\begin{figure}[t]
\includegraphics[width=\columnwidth]{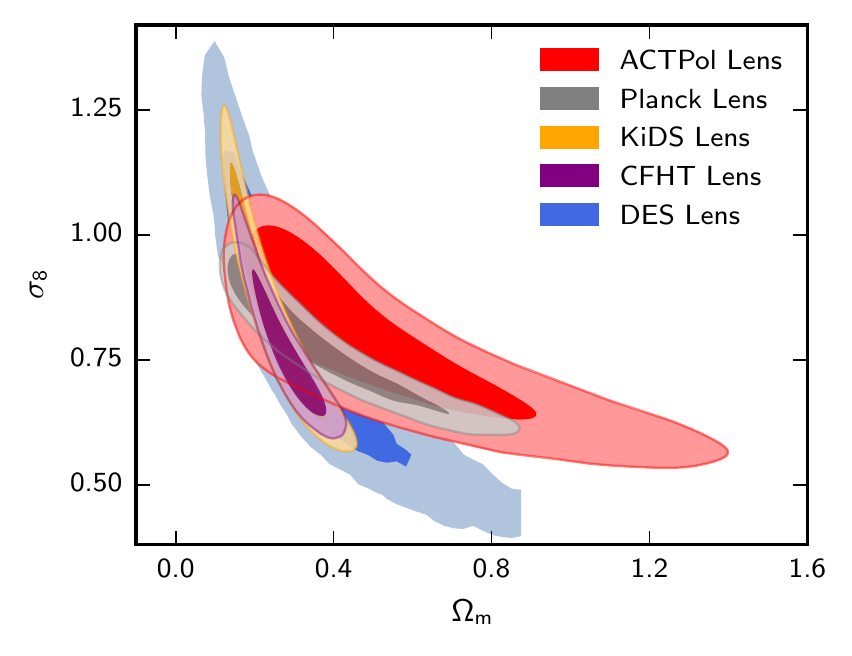}
\caption{Compilation of recent constraints on $\sigma_8$ versus $\Omega_m$ from CMB and optical lensing measurements (CFHTLens \cite{cfht}, KiDS \cite{kidslens} and DES \cite{deslens}). The CMB constraints are from the lensing power spectrum information only.  The datasets are seen to be broadly consistent, and the degeneracy direction from the CMB experiments can be seen to differ from those from the optical surveys.}
\label{fig:LensOnly}
\end{figure}

We present our constraints in Figure~\ref{fig:ACTOnly}.  The red contours show the ACTPol lensing-only results with the source plane fixed in the $R_\phi(L)$ correction to the best-fit \planck\  primary CMB cosmology.  The blue contours show the result when adding BAO, again fixing $\bitheta$ in the $R_\phi(L)$ correction to the \planck\ best-fit model. We compare with the corresponding~\planck\ lensing plus BAO contours shown in grey.  BAO alone has a mild preference for $\Omega_m\approx
0.4$ in this plane, and it intersects the ACTPol only contours around this value.  However, in the $H_0-\Omega_m$ plane, there is only a small parameter region where BAO and \planck\  lens contours intersect, which is around $\Omega_m\approx 0.3$.  Thus, the grey \planck\  lens plus BAO contour is centered around $\Omega_m\approx
0.3$, though the reason for this is not immediately apparent from the $\sigma_8-\Omega_m$ plane alone.  

In Figure~\ref{fig:ACTOnly}, we also show \planck\  primary $TT$ plus BAO and ACTPol primary CMB plus BAO constraints.  CMB power spectrum measurements give a measurement of lensing through peak smearing of the primary spectrum. We call this lensing measurement ``two-point lensing,'' in contrast to the lensing power spectrum measurement discussed in this work, which we call ``four-point lensing.'' 
We note that the \planck\  and ACTPol primary CMB measurements plus BAO are very constraining, both due to their measurements of the two-point lensing signal  and because they constrain the amplitude of high-redshift structure via the optical depth $\tau$.  

In Figure~\ref{fig:LensOnly}, we show a compilation of recent CMB lensing-only (four-point) and optical-lensing only constraints.  The optical lensing constraints are from CFHTLens~\cite{cfht}, KiDS~\cite{kidslens} and DES~\cite{deslens}, and are derived from measurements of galaxy shapes that have been distorted by lensing from intervening matter.  The DES chains provided by the DES team only extend to $\Omega_m \simeq 0.9$.  This plot shows consistency between the data sets given their uncertainties.  

\begin{figure}[t!]
\includegraphics[width=1.0\columnwidth]{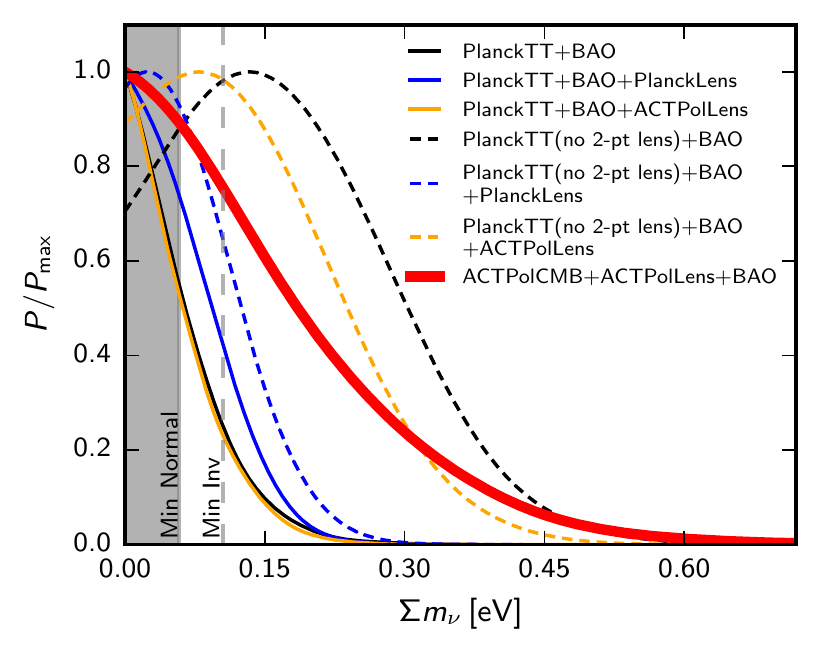}
\caption{Constraint on neutrino mass marginalized over the other parameters from ACTPol primary CMB + ACTPol four-point lensing + BAO (thick red curve).  We also show \planck\  $TT$ + BAO (black solid curve), \planck\  $TT$ + BAO + \planck\  four-point lens (blue solid curve), and \planck\  $TT$ + BAO + ACTPol four-point lens (gold solid curve).  The black dashed curve shows \planck\  $TT$ + BAO when we remove the two-point lensing signal from peak smearing in \planck\  $TT$ (see details in text), and the gold dashed shows the result when ACTPol four-point lensing is added to that.  The difference between black and gold dashed curves isolates the improvement when adding ACTPol four-point lensing data, and comparison of gold dashed and solid curves shows the effect from adding the two-point lensing information in \planck\  $TT$.  We also show the minimal neutrino masses for the normal and inverted hierarchy of 58 and 105 meV, respectively, assuming that the cosmological neutrinos have the same properties as those measured in terrestrial experiments \cite{Agashe:2014kda}.}
\label{fig:mnu}
\end{figure}

To constrain the sum of neutrino masses, we combine our lensing measurement with the ACTPol two-season CMB temperature and polarization power spectra~\cite{louis}, and with BAO. Since our lensing maps are nearly noise-dominated and since we use a data-dependent Gaussian bias subtraction, we can neglect the covariance of the lensing and CMB power spectrum measurements~\cite{schmittfull}. With this combination, we obtain a constraint of

%\be
\begin{align}	
\begin{split}
\Sigma m_{\nu} < \mnuConstraint \mathrm{~eV} \quad &(\text{ACTPol lens}\\
&\text{+ACTPol CMB + BAO, 95\%}).
\label{eq:mnufinal}
\end{split}
\end{align}
%\ee
For this result, the cosmology in the $R_\phi(L)$ correction was allowed to be free.  %We also follow the ``With CMB $TT$'' priors given in Table~\ref{tab:priors}.  
We show this constraint as the thick red curve in Figure~\ref{fig:mnu}.

In Figure~\ref{fig:mnu}, we also show the constraint combining \planck\  four-point lensing plus \planck\  primary CMB $TT$ (the \planck\  temperature power spectrum at all scales) plus BAO, as the solid blue curve. This constraint is $\Sigma m_{\nu} < 0.153 \mathrm{~eV}$ at 95\% CL, which is somewhat tighter than reported in \cite{plancklensing}.  This tightening of the $\Sigma m_{\nu}$ constraint is due to the use of DR12 as opposed to DR11 when including BOSS BAO, and the lower central value and tighter error bar on $\tau$, $0.058\pm0.012$ versus $0.067\pm0.017$, that was recently reported in \cite{newplanckTau}. The constraint without the inclusion of \planck\  four-point lensing is shown as the solid black curve.  

The ACTPol neutrino mass constraint is not yet competitive with those from \planck\ $TT$ and BAO.  Adding ACTPol four-point lensing data leaves this number essentially unchanged, as seen by the solid gold curve in~Figure~\ref{fig:mnu}. (Note that adding the \planck\ four-point lensing instead, the dark blue curve, actually increases the mass limit slightly due to a mild tension between the lensing amplitudes derived from the \planck\ two-point and four-point lensing signals.)

To quantify the constraining power of current ACTPol lensing compared to \planck\  lensing, we freed the parameter $A_\mathrm{lens}$, allowing it to vary just the two-point lensing in the \planck\ $TT$ spectrum \cite{calabrese}.  Marginalizing over this parameter, we effectively removed the lensing information from the \planck\  two-point $TT$ measurement.  We show the result of this as the black dashed curve in Figure~\ref{fig:mnu}, which gives a constraint of $\Sigma m_{\nu} < 0.378 \mathrm{~eV}$ at 95\% CL.  We then added ACTPol four-point lensing, and obtain the gold dashed curve and a constraint of $\Sigma m_{\nu} < 0.320 \mathrm{~eV}$ at 95\% CL.  This improvement is from the ACTPol four-point lensing measurement alone.  The difference with the final constraint given by the solid gold curve shows the weight of the \planck\  two-point lensing signal, which is driven by its high amplitude and tight error bar compared to the \planck\  primary CMB best-fit cosmology.

\section{Conclusions}
We report a new measurement of the power spectrum of CMB lensing from two seasons of ACTPol CMB temperature and polarization data. This measurement can be compared with those of other groups in Fig.~\ref{fig:autoCompilationACTPol}.  We detect lensing power at high significance in our data and find the lensing power spectrum to be consistent with $\Lambda$CDM predictions. No evidence for significant systematic effects is seen in our null tests and checks. We obtain an amplitude of lensing power $\alens = \alensConstraint$, a $\actpolSignificance \sigma$ measurement, and an amplitude of density fluctuations $\sigma_8 = \sigmaEightConstraint$. Both measurements are consistent with the \planck~$\Lambda$CDM cosmology (which we define to have $\alens=1$). While the amplitude of density fluctuations we report is higher than that found in some recent weak lensing surveys~\cite{kidslensing}, our uncertainties are currently still too large to resolve any claimed tensions between \planck\ and these low-redshift tracers. However, we note that our current measurements are based on only 12\% of the ACTPol observational data \cite{louis}. As the remaining ACTPol data are included in our analysis, using the pipeline described in detail in this paper, we expect to report significantly improved measurements of the lensing power spectrum. This will, in turn, give stronger constraints on the amplitude of structure and on cosmological parameters such as the neutrino mass.

\acknowledgements
The authors would like to thank Anthony Challinor, Antony Lewis, and Toshiya Namikawa for useful discussions. This work was supported by the U.S. National Science Foundation through awards AST-1440226, AST- 0965625 and AST-0408698 for the ACT project, as well as awards PHY-1214379 and PHY-0855887. Funding was also provided by Princeton University, the University of Pennsylvania, and a Canada Foundation for Innovation (CFI) award to UBC. ACT operates in the Parque Astron\'omico Atacama in northern Chile under the auspices of the Comisi\'on Nacional de Investigaci\'on Cient\'ifica y Tecnol\'ogica de Chile (CONICYT). Computations were performed on the GPC supercomputer at the SciNet HPC Consortium. SciNet is funded by the CFI under the auspices of Compute Canada, the Government of Ontario, the Ontario Research Fund – Research Excellence; and the University of Toronto. The development of multichroic detectors and lenses was supported by NASA grants NNX13AE56G and NNX14AB58G. NS acknowledges support from NSF grant number 1513618. A.K. has been supported by NSF AST-1312380.  RD and LM thank CONICYT for grants ALMA-CONICYT 31140004, FONDECYT 1141113, Anillo ACT-1417 and BASAL CATA.  We also thank the Mishrahi Fund and the Wilkinson Fund for their generous support of the project.\\

\appendix

\section{The temperature curl null test and the low-$\ell$ cutoff}
\label{sec:appendix}
\begin{figure}[t!]
\includegraphics[width=\columnwidth]{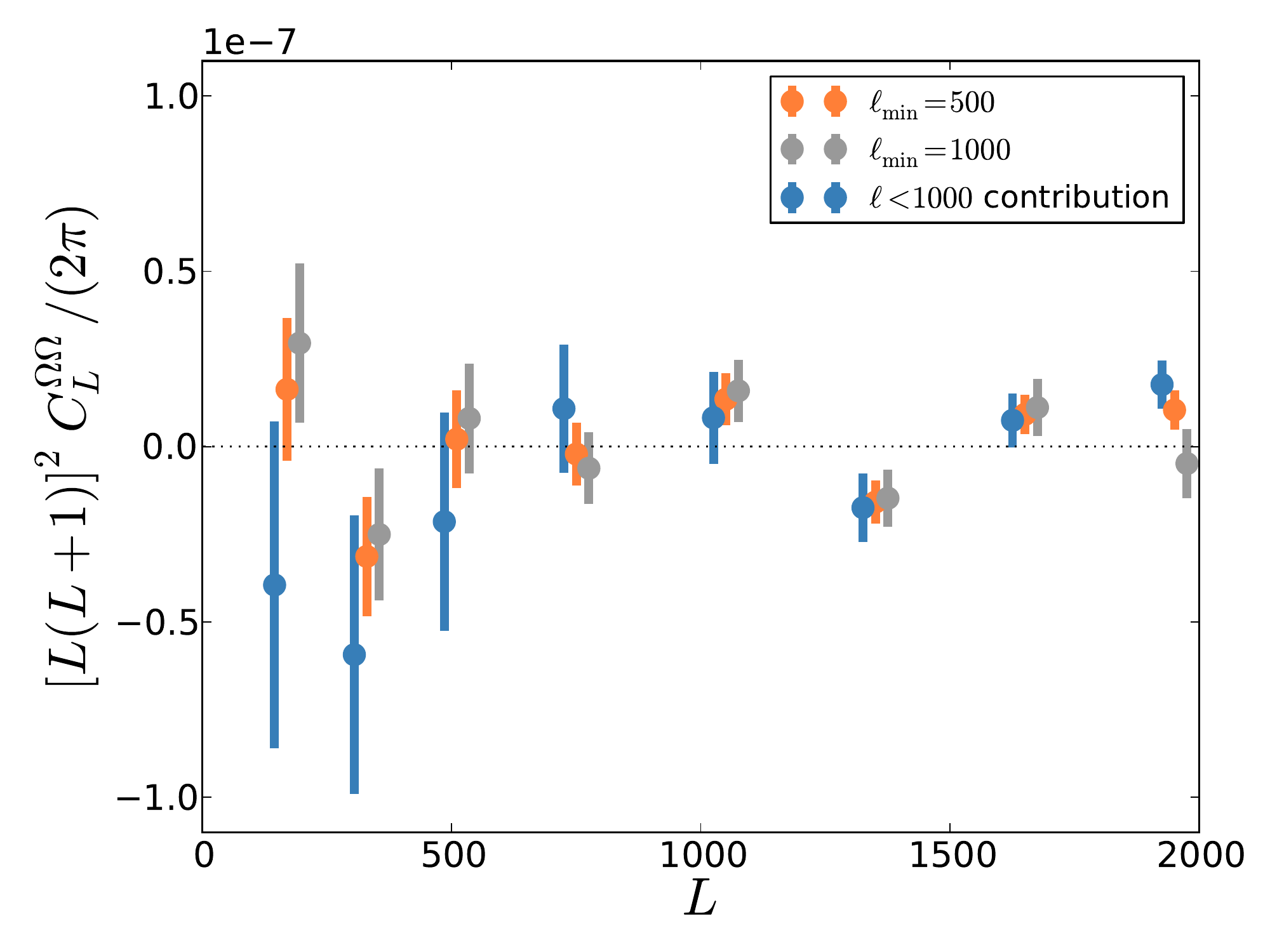}
\label{fig:lowL}
\caption{Curl null test for the D56 $TT,TT$ estimator, with different choices of the low-ell cutoff. It can be seen that the contribution from $\ell<1000$ is in $\sim 3\sigma$  tension with null for the highest $L$ bandpower. This is likely due to the limitations of the realization-dependent bias subtraction, which only corrects for mismatched simulations to leading order, for the highest $L$ bandpower where the bias is very large -- the simulations we use may not adequately capture the noise at $\ell<1000$ where atmospheric noise is very large. The problem is solved by, conservatively, using only CMB scales $\ell>1000$ in our analysis. }
\end{figure}

In our analysis, we impose a lower cutoff on the CMB scales $\ell_{\mathrm{min}}$, from which to measure our lensing. This cut was chosen to be above low multipoles where atmospheric noise is largest. 
Our initial choice was $\ell_{\mathrm{min}}$ = 500. However, with this initial choice, the curl null test for D56 for the $TT,TT$ estimator marginally fails at the 3$\sigma$ level (with a PTE of $\sim 0.2\%$). Though with 60 null tests, one failure of this magnitude is not very unlikely (see discussion in the main text of the $TE,TE$ estimator), the D56 $TT,TT$ measurement is particularly important, as it dominates our result. A significant fraction of the tension seemed to arise from the highest $L$ bandpowers, approaching $L=2000$, where the Gaussian bias that we need to subtract off is largest.

As the prescription for calculating the realization dependent bias only self-corrects to first order in differences between simulations and data, we note that a mismatch in the simulated  CMB power at the 10\% level in any region of our map is sufficient to cause failures at the highest L.  One possibility that gives such a mis-simulation is that our noise simulation procedure assumes that the atmospheric noise scales down with the local map weights as white noise, which is not quite true on large scales. We changed our cutoff to $\ell=1000$ in our analyses to address the fact that the bias subtraction might not be precise enough when using this range of scales in the estimator.  This change results in the null test now passing at a slightly better than 2$\sigma$  level, because the highest $L$-bandpower is more consistent with null, though also to some extent because the error bars increased since data was removed.  By varying the new cutoff above $\ell=1000$, we verified the stability of the result.

To check our understanding and to ensure that the scales we cut are at least partially responsible for the marginal null failure beyond merely inflating error bars, we plot the relevant null test points in Figure~10 for both $\ell_{\mathrm{min}}=500$ and $\ell_{\mathrm{min}}=1000$. We now seek to approximately isolate the new information arising from the low $\ell$ scales by assuming an independent measurement which is coadded with the $\ell_{\mathrm{min}}=1000$ data to obtain the $\ell_{\mathrm{min}}=500$ data. We invert the simple coadd procedure to obtain a new null test, which is also shown on the plot. We approximately identify this null with the contribution that we are cutting, i.e.~the part that originates below $\ell_{\mathrm{min}}=1000$ (noting that in noise domination with a realization dependent bias, the correlation of the four-point functions involving any $\ell<1000$ contribution with the $\ell_{\mathrm{min}}=1000$ measurement is small.) 

It can be seen that the highest-$L$ bandpower deviates at the $3\sigma$ level from null for the $\ell_{\mathrm{min}}<1000$ contribution we isolate. This suggests that the large scales are to some extent responsible for the problem, and is consistent with our picture of mis-simulation of atmospheric noise causing problems in the highest $L$ bandpower where the bias subtraction is largest.  In future work, we plan to prioritize our noise modeling (or alternatively, the development of a cross-spectrum based estimator) to mitigate this issue and extend the range of scales we can use in our analysis.  In addition, with the significant increase in data expected from the full three-season dataset, we will be able to investigate any hints of systematics in our data with more powerful null tests.

\begin{figure*}
\includegraphics[width=7in]{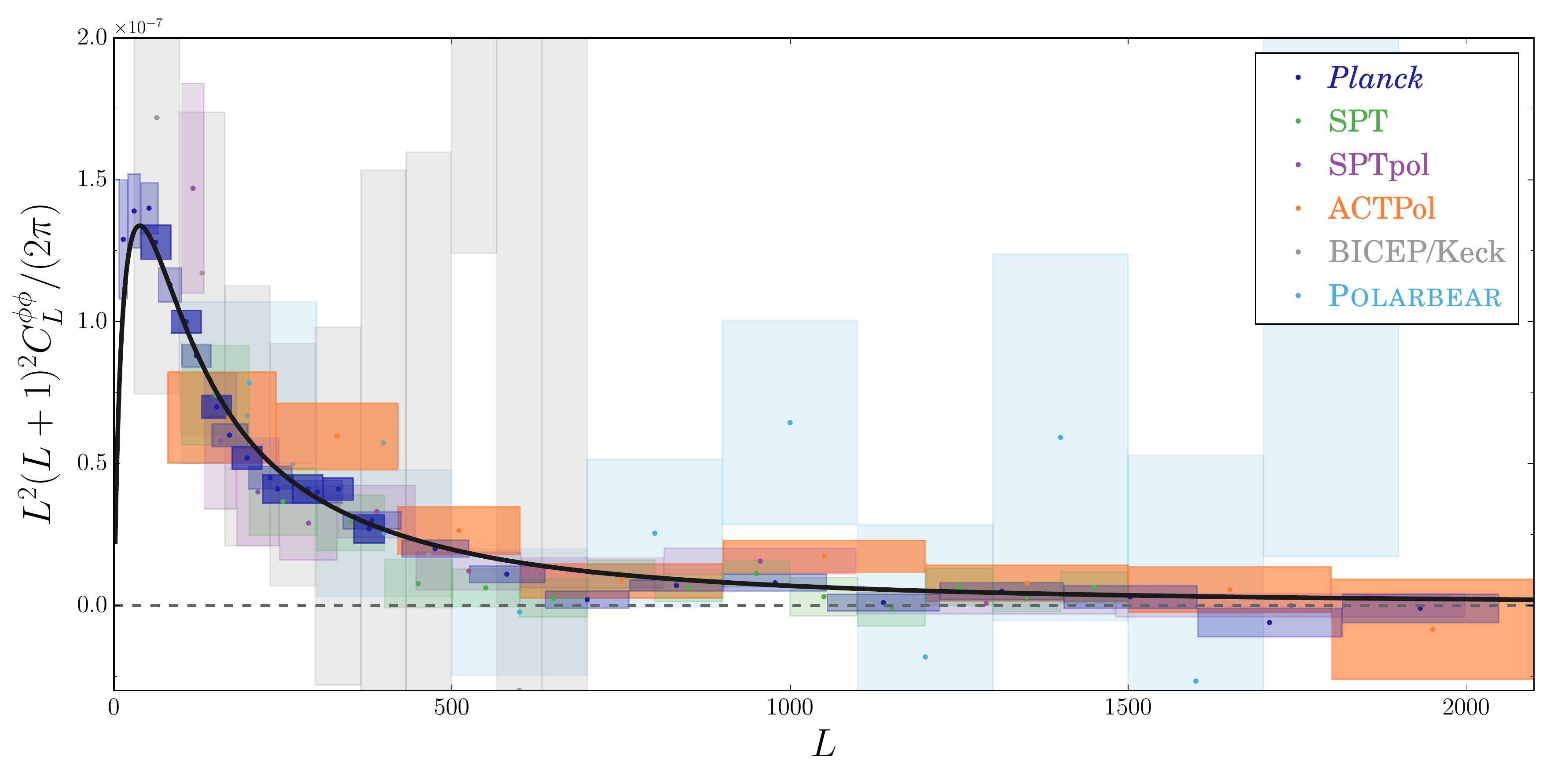}
\caption{Compilation of lensing power spectrum measurements from various experimental groups, including \textit{\planck}\ \cite{plancklensing}, SPT \cite{engelenlensing}, SPTpol \cite{storylensing}, this work, BICEP/Keck \cite{biceplensing}, and Polarbear \cite{polarbearlensingA}.  In the case of \textit{\planck}, we show both the full range of reported bandpowers (light purple), as well as the restricted range $40 < L < 400$ used for cosmological analysis (dark purple).}
\label{fig:autoCompilationACTPol}
\end{figure*}

\end{document}

%% file: latexTableActpol2016.tex
\begin{tabular}{|c | c | c|}
\hline
$L$ & $L^4 C_L^{\phi\phi} / (4 \times 10^{-7})$ & $ \sigma(L^4 C_L^{\phi\phi}) / (4 \times 10^{-7})$ \\
\hline
138 &  1.039 &  0.251 \\
301 &  0.937 &  0.183 \\
484 &  0.414 &  0.132 \\
697 &  0.136 &  0.094 \\
1002 &  0.271 &  0.089 \\
1304 &  0.124 &  0.100 \\
1602 &  0.087 &  0.126 \\
1911 & -0.132 &  0.277 \\
\hline\end{tabular}

%% file: main.bbl
\begin{thebibliography}{3}
	
%intro
\bibitem[Blanchard \& Schneider(1987)]{blanchard87} Blanchard, A., \& Schneider, J.\ 1987, \aap, 184, 1 

\bibitem[Bernardeau(1997)]{bernardeau97} Bernardeau, F.\ 1997, \aap, 324, 15 

\bibitem[Zaldarriaga \& Seljak(1999)]{zaldarriaga98} Zaldarriaga, M., \& Seljak, U.\ 1999, \prd, 59, 123507 

\bibitem[Lewis \& Challinor(2006)]{lewischallinor} Lewis, A., \& Challinor, A.\ 2006, \physrep, 429, 1 

\bibitem[Natarajan et al.(2014)]{natarajan} Natarajan, A., Zentner, A.~R., Battaglia, N., \& Trac, H.\ 2014, \prd, 90, 063516 

\bibitem[Namikawa(2016)]{namikawa16} Namikawa, T.\ 2016, \prd, 93, 121301 

\bibitem[Liu et al.(2016)]{liu} Liu, J., Hill, J.~C., Sherwin, B.~D., et al.\ 2016, \prd, 94, 103501 

\bibitem[B{\"o}hm et al.(2016)]{boehm} B{\"o}hm, V., Schmittfull, M., \& Sherwin, B.~D.\ 2016, \prd, 94, 043519 



\bibitem[Kosowsky(2003)]{kosowskyatacama} Kosowsky, A.\ 2003, New Astronomy Reviews, 47, 939 


\bibitem[Ruhl et al.(2004)]{ruhlspt} Ruhl, J., Ade, P.~A.~R., Carlstrom, J.~E., et al.\ 2004, \procspie, 5498, 11 


\bibitem[The Planck Collaboration(2006)]{planckbluebook} The Planck Collaboration 2006, arXiv:astro-ph/0604069 



\bibitem[Smith et al.(2007)]{smithlensing} Smith, K.~M., Zahn, O., \& Dor{\'e}, O.\ 2007, \prd, 76, 043510 

\bibitem[Hirata et al.(2008)]{hiratalensing} Hirata, C.~M., Ho, S., Padmanabhan, N., Seljak, U., \& Bahcall, N.~A.\ 2008, \prd, 78, 043520 

\bibitem[Das et al.(2011)]{actlensing} Das, S., Sherwin, B.~D., Aguirre, P., et al.\ 2011, Physical Review Letters, 107, 021301 

\bibitem[Sherwin et al.(2011)]{actparams} Sherwin, B.~D., Dunkley, J., Das, S., et al.\ 2011, Physical Review Letters, 107, 021302 

\bibitem[van Engelen et al.(2012)]{engelenlensing} van Engelen, A., Keisler, R., Zahn, O., et al.\ 2012, \apj, 756, 142 

\bibitem[Ade et al.(2014)]{polarbearlensingA} Ade, P.~A.~R., Akiba, Y., Anthony, A.~E., et al.\ 2014, Physical Review Letters, 113, 021301 

\bibitem[Ade et al.(2014)]{polarbearlensingB} Ade, P.~A.~R., Akiba, Y., Anthony, A.~E., et al.\ 2014, Physical Review Letters, 112, 131302 

\bibitem[Hanson et al.(2013)]{hansonBmode} Hanson, D., Hoover, S., Crites, A., et al.\ 2013, Physical Review Letters, 111, 141301 

\bibitem[Story et al.(2015)]{storylensing} Story, K.~T., Hanson, D., Ade, P.~A.~R., et al.\ 2015, \apj, 810, 50 

\bibitem[Keck Array et al.(2016)]{biceplensing} Keck Array, T., BICEP2 Collaborations, :, et al.\ 2016, arXiv:1606.01968 

\bibitem[\planck\ Collaboration et al.(2014)]{plancklensing2013} \planck\ Collaboration, Ade, P.~A.~R., Aghanim, N., et al.\ 2014, \aap, 571, A17 

\bibitem[Planck Collaboration et al.(2016)]{plancklensing} Planck Collaboration, Ade, P.~A.~R., Aghanim, N., et al.\ 2016, \aap, 594, A15 

\bibitem[Hildebrandt et al.(2016)]{kidslensing} Hildebrandt, H., Viola, M., Heymans, C., et al.\ 2016, arXiv:1606.05338 

\bibitem[De Bernardis et al.(2016)]{ACTPolSurveyStrategy} De Bernardis, F., Stevens, J.~R., Hasselfield, M., et al.\ 2016, arXiv:1607.02120 

\bibitem[van Engelen et al.(2015)]{engelencib} van Engelen, A., Sherwin, B.~D., Sehgal, N., et al.\ 2015, \apj, 808, 7 

\bibitem[Allison et al.(2015)]{allison} Allison, R., Lindsay, S.~N., Sherwin, B.~D., et al.\ 2015, \mnras, 451, 849 

\bibitem[Madhavacheril et al.(2015)]{madhavacherillensing} Madhavacheril, M., Sehgal, N., et al.\ 2015, Physical Review Letters, 114, 151302 

\bibitem[Thornton et al.(2016)]{actpolinstrument} Thornton, R.~J., Ade, P.~A.~R., Aiola, S., et al.\ 2016, arXiv:1605.06569 

\bibitem[Louis et al.(2016)]{louis} Louis, T., Grace, E., Hasselfield, M., et al.\ 2016, arXiv:1610.02360 

\bibitem[Naess et al.(2014)]{naess} Naess, S., Hasselfield, M., McMahon, J., et al.\ 2014, \jcap, 10, 007 

\bibitem[Bucher \& Louis(2012)]{bucherlouis} Bucher, M., \& Louis, T.\ 2012, \mnras, 424, 1694 

\bibitem[Smith(2006)]{smithB} Smith, K.~M.\ 2006, \prd, 74, 083002 

\bibitem[Pearson et al.(2014)]{pearsonlens} Pearson, R., Sherwin, B., \& Lewis, A.\ 2014, \prd, 90, 023539 

\bibitem[Calabrese et al.(2013)]{calabreseparams} Calabrese, E., Hlozek, R.~A., Battaglia, N., et al.\ 2013, \prd, 87, 103012 

\bibitem[Louis et al.(2013)]{louislensing} Louis, T., N{\ae}ss, S., Das, S., Dunkley, J., \& Sherwin, B.\ 2013, \mnras, 435, 2040 

\bibitem[Namikawa et al.(2013)]{namikawabias} Namikawa, T., Hanson, D., \& Takahashi, R.\ 2013, \mnras, 431, 609 

\bibitem[Hu \& Okamoto(2002)]{huokamoto} Hu, W., \& Okamoto, T.\ 2002, \apj, 574, 566 

\bibitem[Calabrese et al.(2008)]{calabrese} Calabrese, E., Slosar, A., Melchiorri, A., Smoot, G.~F., \& Zahn, O.\ 2008, \prd, 77, 123531 


\bibitem[Hanson et al.(2011)]{hanson2010} Hanson, D., Challinor, A., Efstathiou, G., \& Bielewicz, P.\ 2011, \prd, 83, 043005 

\bibitem[Schmittfull et al.(2013)]{schmittfull} Schmittfull, M.~M., Challinor, A., Hanson, D., \& Lewis, A.\ 2013, \prd, 88, 063012 

\bibitem[Green et al.(2016)]{green} Green, D., Meyers, J., \& van Engelen, A.\ 2016, arXiv:1609.08143 


\bibitem[Peloton et al.(2016)]{peloton} Peloton, J., Schmittfull, M., Lewis, A., Carron, J., \& Zahn, O.\ 2016, arXiv:1611.01446 




\bibitem[Pratten \& Lewis(2016)]{prattenlewis} Pratten, G., \& Lewis, A.\ 2016, \jcap, 8, 047 

\bibitem[Marozzi et al.(2016)]{marozzi} Marozzi, G., Fanizza, G., Di Dio, E., \& Durrer, R.\ 2016, \jcap, 9, 028 

\bibitem[Osborne et al.(2014)]{osbornebias} Osborne, S.~J., Hanson, D., \& Dor{\'e}, O.\ 2014, \jcap, 3, 024 

\bibitem[van Engelen et al.(2014)]{engelenbias} van Engelen, A., Bhattacharya, S., Sehgal, N., et al.\ 2014, \apj, 786, 13 

\bibitem[Sehgal et al.(2010)]{sehgalsims} Sehgal, N., Bode, P., Das, S., et al.\ 2010, \apj, 709, 920 

\bibitem[Hartlap et al.(2007)]{hartlap} Hartlap, J., Simon, P., \& Schneider, P.\ 2007, \aap, 464, 399 


\bibitem[Smith et al.(2003)]{halofit2003} Smith, R.~E., Peacock, J.~A., Jenkins, A., et al.\ 2003, \mnras, 341, 1311 

\bibitem[Takahashi et al.(2012)]{halofit2012} Takahashi, R., Sato, M., Nishimichi, T., Taruya, A., \& Oguri, M.\ 2012, \apj, 761, 152 

\bibitem[Lewis(2013)]{cosmomc1} Lewis, A.\ 2013, \prd, 87, 103529 

\bibitem[Lewis \& Bridle(2002)]{cosmomc2} Lewis, A., \& Bridle, S.\ 2002, \prd, 66, 103511 

\bibitem[Planck Collaboration (2016)]{newplanckTau} \planck\ Collaboration, Adam, R., Aghanim, N., et al.\ 2016, arXiv:1605.03507 

\bibitem[Pettini \& Cooke(2012)]{PettiniCooke} Pettini, M., \& Cooke, R.\ 2012, \mnras, 425, 2477 

\bibitem[Planck Collaboration (2016)]{planckParams2015} \planck\ Collaboration, Ade, P.~A.~R., Aghanim, N., et al.\ 2016, \aap, 594, A13 

\bibitem[Beutler et al.(2011)]{sixdf} Beutler, F., Blake, C., Colless, M., et al.\ 2011, \mnras, 416, 3017 

\bibitem[Ross et al.(2015)]{mgs} Ross, A.~J., Samushia, L., Howlett, C., et al.\ 2015, \mnras, 449, 835 

\bibitem[Gil-Mar{\'{\i}}n et al.(2016)]{bossdr12} Gil-Mar{\'{\i}}n, H., Percival, W.~J., Cuesta, A.~J., et al.\ 2016, \mnras, 460, 4210 
%\bibitem[Beutler et al.(2016)]{Beutler} Beutler, F., Seo, H.-J., Ross, A.~J., et al.\ 2016, \mnras, 


\bibitem[Joudaki et al.(2016)]{cfht} Joudaki, S., Blake, C., Heymans, C., et al.\ 2016, arXiv:1601.05786 

\bibitem[Hildebrandt et al.(2016)]{kidslens} Hildebrandt, H., Viola, M., Heymans, C., et al.\ 2016, arXiv:1606.05338 

\bibitem[The Dark Energy Survey Collaboration et al.(2015)]{deslens} The Dark Energy Survey Collaboration, Abbott, T., Abdalla, F.~B., et al.\ 2015, arXiv:1507.05552 

%\cite{Agashe:2014kda}
\bibitem{Agashe:2014kda} 
  K.~A.~Olive { et al.} 2015 [Particle Data Group Collaboration],
  %``Review of Particle Physics,''
  Chin.\ Phys.\ C {38}, 090001

  %%CITATION = doi:10.1088/1674-1137/38/9/090001;%%
  %5349 citations counted in INSPIRE as of 28 Nov 2016



\end{thebibliography}
